\documentclass{aa}  
\usepackage{graphicx}
\usepackage{natbib}
\usepackage{verbatim}
\usepackage{pdfpages}
\usepackage{caption}
\bibpunct{(}{)}{;}{a}{}{,}

\usepackage{txfonts}
\begin{document}
   \title{Identification of new transitional disk candidates in Lupus with \textit{Herschel} \thanks{ \textit{Herschel} is an ESA space observatory with science instruments provided by European-led Principal Investigator consortia and with important participation from NASA.}}
     
   \author{I. Bustamante \inst{1,2,3}
   \and B. Mer\'in \inst{1}
   \and \'A. Ribas \inst{1,2,3}
   \and H. Bouy \inst{2}
   \and T. Prusti \inst{4}
   \and G. L. Pilbratt \inst{4}
   \and Ph. Andr\'e \inst{5}
               }

   \institute{European Space Astronomy Centre (ESA), P.O. Box, 78, 28691 Villanueva de la Ca\~{n}ada, Madrid, Spain
   \and Centro de Astrobiolog\'ia, INTA-CSIC, P.O. Box - Apdo. de correos 78, Villanueva de la Ca\~{n}ada Madrid 28691, Spain
   \and ISDEFE - ESAC, P.O. Box, 78, 28691 Villanueva de la Ca\~{n}ada, Madrid, Spain
   \and ESA Science Support Office, ESTEC/SRE-S, Keplerlaan 1, 2201 AZ Noordwijk, The Netherlands
   \and Laboratoire AIM Paris, Saclay, CEA/DSM, CNRS, Universit\'e Paris Diderot, IRFU, Service d'Astrophysique, Centre d'Etudes de Saclay, Orme des Merisiers, 91191 Gif-sur-Yvette, France
   }
      \date{Accepted by A\&A on January 15th, 2015}

% \abstract{}{}{}{}{} 
% 5 {} token are mandatory
 
  \abstract
% context heading (optional)
  % {} leave it empty if necessary  
   {New data from the \textit{Herschel} Space Observatory are broadening our understanding of the physics and evolution of the outer regions of protoplanetary disks in star forming regions. In particular they prove to be useful to identify transitional disk candidates.}
  % aims heading (mandatory)
   {The goals of this work are to complement the detections of disks and the identification of transitional disk candidates in the Lupus clouds with data from the \textit{Herschel} Gould Belt Survey.}
  % methods heading (mandatory)
   {We extracted photometry at 70, 100, 160, 250, 350 and 500 $\mu$m of all spectroscopically confirmed Class II members previously identified in the Lupus regions and analyzed their updated spectral energy distributions.}
  % results heading (mandatory) 
  {We have detected 34 young disks in Lupus in at least one \textit{Herschel} band, from an initial sample of 123 known members in the observed fields. Using the criteria defined in Ribas et al. (2013) we have identified five transitional disk candidates in the region. Three of them are new to the literature. Their PACS-70 $\mu$m fluxes are systematically higher than those of normal T Tauri stars in the same associations, as already found in T Cha and in the transitional disks in the Chamaeleon molecular cloud.}
  % conclusions heading (optional), leave it empty if necessary 
{\textit{Herschel} efficiently complements mid-infrared surveys for identifying transitional disk candidates and confirms that these objects seem to have substantially different outer disks than the T Tauri stars in the same molecular clouds. }


 \maketitle
%
%________________________________________________________________

\section{Introduction}

Transitional disks link the studies of planet formation and disk evolution around young stars. They are protoplanetary disks around young stars, optically thick and gas rich, with astronomical unit-scale inner disk clearings or radial gaps evident in their Spectral Energy Distributions (hereafter, SED). They lack excess emission at short-mid infrared (typically around 8-12 $\mu$m) but present considerable emission at longer wavelengths \citep{Muzerolle2010}. Since their discovery by \cite{Strom1989} with IRAS data, the \textit{Spitzer} Space Telescope \citep{Werner2004}  allowed the identification of a much larger population of these objects \citep{Cieza2010, Merin2010, Espaillat2014}. It was possible to use the mid-infrared data to make good estimates of the dust distribution in the inner few astronomical units of the disks around T Tauri stars in nearby star-forming regions \citep[e.g.][]{Calvet2005,Lada2006}. 

The \textit{Herschel} Space Observatory \citep{Pilbratt2010} grants us access to the outer disk region of these objects. Recent studies suggest that the outer disks, as detected in far-infrared emission, undergo substantial transformations in the transitional disk phase \citep{Cieza2011, Ribas2013}. However, \textit{Herschel} data are still lacking for most known transitional disks.

The main objective of this article is to revisit and extend our knowledge of known Class II objects in the Lupus dark clouds, and to identify transitional disk candidates in the region using \textit{Herschel} photometry, complemented with photometry from previous studies from the optical to the mid-infrared. With \textit{Herschel} we have the opportunity to study objects with large inner holes, reaching for the furthest regions of their disks, where \textit{Spitzer} was unable to access with the required sensitivity (IRAC and IRS had different wavelength ranges, between 3.5 and 8.0 $\mu$m and 5 and 35 $\mu$m, respectively, and the sensitivity of MIPS, with pass bands between 24 and 160 $\mu$m, was limited). Therefore, we are not only able to study the outer regions of known transitional disks, but also detect new candidates, with larger cavities and outer disks.

This work is organized as follows: \S~\ref{observations} describes the \textit{Herschel} observations and data reduction and the procedure applied to identify and extract the fluxes for the known objects. In \S~\ref{results} we identify new transitional disk candidates,  describe the procedure to build the SEDs and discuss some objects with certain peculiarities. In \S~\ref{discussion} we discuss the properties of the newly found transitional disk candidates and in \S~\ref{conclusions} we present the conclusions of this work.

\section{Observations and data reduction}
\label{observations}

The Lupus dark clouds were observed by \textit{Herschel} as part of the \textit{Herschel} Gould Belt Survey \citep[HGBS hereafter, ][]{Andre2010}. It is one of the star-forming regions located within 150 to 200 pc from the Sun and has a mean age of a few Myrs \cite[see ][for a review about the region]{Comeron2008}. It contains four main regions, with one of
them, Lupus III, being a rich T Tauri association. Located in the Scorpius-Centaurus OB association, the massive stars in the region are likely to have played a significant role in its evolution.

By employing the PACS \citep{Poglitsch2010} and SPIRE \citep{Griffin2010} photometers in parallel mode, \textit{Herschel} has the capability for large scale mapping of star forming regions in the far-infrared range. Early \textit{Herschel} observations of Lupus from the HGBS were already published by \cite{Rygl2013}, where the observing strategy and main results are described.

We produced our maps using \textit{Scanamorphos} \citep{Roussel2012} version 21.0 for the PACS maps and version 18.0 for the SPIRE ones. The obsids of the Lupus \textit{Herschel} parallel mode observations processed for this work are (1342-)189880, 189881, 203070, 2203071, 2203087, 2203088, 2213182 and 2213183 from program KPGT\_pandre\_1.
The observations were made in parallel mode for 70, 160, 250, 350 and 500 $\mu$m, and in prime mode for 100 $\mu$m \citep{Rygl2013}. This is the reason why some of the sources do not have images at 100 $\mu$m in Figures \ref{fig:images} to \ref{fig:images5}. 

A complete census of the sources detected with \textit{Herschel} will be published by the \textit{Herschel} Gould Belt Survey consortium in an upcoming paper (Benedettini et al. in prep.).

\subsection{Point source photometry}
\label{fluxes}
Our first objective was to detect and classify sources in the images obtained with PACS and SPIRE. We used coordinates of spectroscopically confirmed Class II members from different sources, compiling a list of 217 targets. This compilation was developed by \cite{Ribas2014}, where these objects were selected for having known extinction values and spectral types from optical spectroscopy. Of this sample of 217 objects, only 123 are found in the fields observed by \textit{Herschel}. 

We applied the \textit{sourceExtractorSussextractor} algorithm in our maps, with the \textit{Herschel Interactive Processing Environment} (HIPE), version 12.0, using a detection threshold of S/N${>}$ 3. Then we cross-matched the resulting list of detected objects with our pre-selected targets, using a search radius between 5" and 10". We selected only those objects with detections in PACS. After this process, we found 37 objects with at least one detection in one PACS band, 35 in the 70 $\mu$m maps and 2 in the 100 $\mu$m maps.
We then validated all of them by visual inspection. This way we were able to double check our detections. The background emission becomes more significant at longer wavelengths, where detections are in general less frequent, or false detections appear, due to bright filaments and ridges. This is the reason why we selected only PACS-detected objects. The combination of quantitative and qualitative analysis was essential to produce reliable detections and photometry.  After this process we discarded 4 objects as not having clear detections in the maps and added 2 objects that were found by visual inspection. They are flagged in Table \ref{tab:stellarparameters}.

After this process, 35 objects of the list had one clean detection in at least one \textit{Herschel} PACS or SPIRE band.

We applied the \textit{sourceExtractorDaophot} algorithm to extract the photometry for PACS, and the \textit{annularSkyAperturePhotometry} task for SPIRE. All point-source fluxes were aperture corrected with the recommended aperture corrections for both instruments (see the PACS Photometer Point Spread Function Technical Note from April 2012, and Sect. 6.7.1 of the SPIRE Data Reduction Guide of HIPE 12.0). Table \ref{tab:aperturePhotometry} shows the set of aperture, inner and outer sky radii for each \textit{Herschel} band, as well as the corresponding aperture corrections.

\begin{table}[htbp]
\caption{Aperture, inner and outer radii used in the photometry extraction process for each band, with their corresponding aperture corrections.}
\centering
\begin{tabular}[width=0.5\textwidth]{lc c c c cl}
\hline\hline
Band & R$_{aperture}$ ('') & R$_{inner}$ ('') & R$_{outer}$ ('') & Correction\\
\hline
PACS70 & 6 & 25&35&1.5711\\
PACS100 & 6 & 25&35&1.6804\\
PACS160 & 12 & 25&35&1.4850\\
SPIRE250 & 22 & 60&90&1.2796\\
SPIRE350 & 30 &60&90&1.2396\\
SPIRE500 & 42 & 60&90&1.2937\\
\hline
\label{tab:aperturePhotometry}
\end{tabular}
\end{table}

Table \ref{tab:sensitivities} shows the sensitivity limit for each \textit{Herschel} band, as stated in the SPIRE/PACS Parallel Mode Observers' Manual (section 2.3). The values correspond to a nominal scan direction and a scan speed of 60"/s.

\begin{table}[htbp]
\caption{Sensitivity limits for each \textit{Herschel} band in parallel mode at 60"/s, as stated in the PACS/SPIRE Parallel Mode Observers' Manual.}
\centering
\begin{tabular}[width=0.5\textwidth]{lc cl}
\hline\hline
Filter & Sensitivity (mJy/beam)\\
\hline
PACS 70 & 21.0 \\
PACS 100 & 24.7 \\
PACS 160 & 47.0 \\
SPIRE 250 & 12.6 \\
SPIRE 350 & 10.5 \\
SPIRE 500 & 15.0\\
\hline
\label{tab:sensitivities}
\end{tabular}
\end{table}

Upper limits for non-detections were derived by adding scaled synthetic Point Spread Functions (PSFs) at the expected position of the sources until the \textit{sussExtractor} detection algorithm did not recover them. In most cases visual inspection was applied. The calibration errors for PACS and SPIRE are 5\% and 7\% (as stated in their observer manuals), respectively. To keep consistency with Ribas et al. (2013) we chose a more conservative estimation, 20\% for SPIRE as error values, taking into account that the background emission becomes increasingly stronger at longer wavelengths. In the case of PACS, after comparing several photometric extraction algorithms (\textit{sussExtractor}, \textit{Daophot}, \textit{annularSkyAperturePhotometry} and \textit{Hyper} \citep{Traficante2014}), and extraction apertures with the MIPS70 fluxes of a subsample of clean sources, we selected a conservatively error value of 25\% of the flux, to account for the different flux values obtained. Table \ref{tab:stellarparameters} gives the stellar parameters of these detected sources, as given in the literature, and Table \ref{tab:table2} presents their \textit{Herschel} point source fluxes and upper limits. As stated before, at least one point source was detected for at least one wavelength of \textit{Herschel}’s instruments for all objects in Table \ref{tab:table2} (see Figures 5 to 9).

We then looked for bright \textit{Spitzer} 24 $\mu$m and 70 $\mu$m sources around the targets, to check for possible contamination from other nearby mid-infrared sources. We used a radius of 42", as this was the value we used for the largest wavelength band when extracting the \textit{Herschel} photometry (which corresponds to SPIRE 500 $\mu$m). The results of this analysis are shown in Tables \ref{tab:counterpartsMIPS24} and \ref{tab:counterpartsMIPS70}.

In the cases where MIPS70 sources were found near our sources we inspected the PACS70 images of the targets. Except for sources Sz65, Sz66, Sz103, Sz104 and 2MASSJ1608537 (as explained below), we did not found evidence of the presence of extra sources inside the matching radius (see Figures \ref{fig:images} to \ref{fig:images5}). This means that \textit{Herschel} did not detect them, so we assume that the full contribution to the flux measured at PACS70 comes from our target sources, and not from contaminants. In Table \ref{tab:counterpartsMIPS70} we maintain all the sources with possible contamination for reference.

Objects Sz65 and Sz66 form a pair, as is explained in Section \ref{Sz65Sz66}, and this is taken into account in their photometry extraction. The MIPS excesses seen for objects Sz103 are related with object Sz104, and vice versa. We only flagged objects with potential contaminations in case they were detected (see Table \ref{tab:table2}). When they are not detected and upper limits are calculated instead, we do not take into account the possible contaminations.

The nearby source near object 2MASSJ1608537 presented significantly larger MIPS24 and MIPS70 fluxes than our target. In this case, we concluded that the infrared excesses measured in the \textit{Herschel} image corresponded to this other object instead of our target, and thus discarded object 2MASSJ1608537, as its \textit{Herschel} photometry could not be unequivocally associated with it. The other object is a known YSO candidate, SSTc2d J160853.2-391440. It is probably another member of the Lupus region, yet not confirmed spectroscopically and therefore not included in our study.

In other cases other contaminant sources were found, but with sufficiently weaker MIPS24 and MIPS70 fluxes as not to affect our measurements. These sources are marked in Table \ref{tab:counterpartsMIPS24}. In these cases we assumed that the \textit{Herschel} fluxes measured correspond to our targets as the major contributors.

Another interesting object is Sz108B, which presents two possible contaminants that could affect its photometry. The one furthest away, at 38.8", has larger MIPS24 flux. The closest one, at 9.7", has a higher MIPS70 flux. Thus, the \textit{Herschel} photometry of Sz108B could be contaminated by the closest one and therefore it is flagged as unreliable in Table \ref{tab:counterpartsMIPS70}.

After this analysis, our final set of sources decreased to 34. We maintain the image of object 2MASSJ1608537 in Figure \ref{fig:images3} for completeness.

\begin{table*}[ht]
\centering
\caption{Targets with possible contamination within the used apertures in MIPS24. The second column shows the fraction of total MIPS24 flux in the aperture with that of the target.}
\begin{tabular}[width=1.0\textwidth]{lc c c c c c c c cl}
\hline\hline
Object &  F$_{\rm source}$/F$_{\rm aper}$ & N  & Companion distance (") & Band \\
\hline
Sz65\tablefootmark{1}  & 0.4&1&6.5&PACS70\\
Sz66\tablefootmark{1}  & 2.9&1&6.2&PACS70   \\
Sz100\tablefootmark{3} &  0.2&1&23.4&SPIRE350   \\
Sz103\tablefootmark{2}  & 0.8 &10&41.2&SPIRE500\\
Sz104\tablefootmark{2}  & 2.7&10&36.4&SPIRE500\\
Sz108B & 1.4&3&38.8&SPIRE500 \\
2MASSJ1608537\tablefootmark{*}& 9.3 &1&6.9& PACS70 \\
Sz112 & 0.5&4&31.9&SPIRE500\\
SSTc2dJ161029.4-392215\tablefootmark{3}& 0.1&2&7.4&PACS70\\
SSTc2dJ160002.4-422216\tablefootmark{3}& 0.4&2&39.8&SPIRE500\\
SSTc2dJ160111.6-413730\tablefootmark{3}&0.2&1&24.2&SPIRE350\\
\hline
\label{tab:counterpartsMIPS24}
\end{tabular}
\tablefoot{
\tablefootmark{1}{The MIPS24 contamination affecting Sz65 is caused by Sz66, and vice versa, as can be seen in Figure \ref{fig:images}. This is taken into account in the photometry analysis.} \\
\tablefootmark{2}{The MIPS24 contamination affecting Sz103 is caused by Sz104, and vice versa, as can be seen in Figures \ref{fig:images2} and \ref{fig:images3}. This is taken into account in the photometry analysis.} \\
\tablefootmark{3}{Objects with sufficiently low contamination from nearby sources as to not affect our \textit{Herschel} measurements.}\\
\tablefootmark{*}{This is the only object that has been removed from the list of detected sources, because its \textit{Herschel} photometry cannot be unequivocally associated to it.}\\
}
\end{table*}

\begin{table*}[ht]
\centering
\caption{Same as Table \ref{tab:counterpartsMIPS24}, but with MIPS70 possible contamination.}
\begin{tabular}[width=1.0\textwidth]{lc c c c c c c c cl}
\hline\hline
Object &  F$_{\rm source}$/F$_{\rm aper}$ & N& Companion distance (") & Band\\
\hline
Sz66  & 468.\tablefootmark{mJy}&1&6.2&PACS70\\
Sz98\tablefootmark{2} &0.2 &1&26.3&SPIRE350  \\
Sz100\tablefootmark{2} &0.7 &1&32.3&SPIRE500   \\
Sz103\tablefootmark{1}  &1.4&1&23.2&SPIRE350\\
Sz104\tablefootmark{1}  & 0.7&1&23.2&SPIRE350\\
Sz108B &212.\tablefootmark{mJy}&1&9.7&PACS70 \\
2MASSJ1608537\tablefootmark{*}&119.0\tablefootmark{mJy}&1&6.9&PACS70\\
Sz112\tablefootmark{2}&0.7 &1&28.6&SPIRE350\\
\hline
\label{tab:counterpartsMIPS70}
\end{tabular}
\tablefoot{
\tablefootmark{1}{The MIPS70 contamination in Sz103 is caused by Sz104, and vice versa, as can be seen in Figure \ref{fig:images2} and \ref{fig:images3}. This is taken into account in the photometry analysis.} \\
\tablefootmark{2}{Objects with sufficiently low contamination from nearby sources as to not affect our \textit{Herschel} measurements.}\\
\tablefootmark{*}{This is the only object that has been removed from the list of detected sources, because its \textit{Herschel} photometry cannot be unequivocally associated to it.}\\
\tablefootmark{mJy}{In these cases the MIPS photometry of the target was not available, but the one of the companions was. Therefore, we present the actual flux of the objects inside the search radius, instead of the percentage.} 
}
\end{table*}

\section{Results}
\label{results}
\subsection{Identification of transitional disk candidates}
\label{trans_disk_identification}
To identify transitional disk candidates we applied the procedure developed in \cite{Ribas2013}, which defines a transitional disk candidate as having $\alpha$$_{K-12}$ ${<}$ 0 and
$\alpha$$_{12-70}$ ${>}$ 0, where $\alpha$$_{\lambda_{1}-\lambda_{2}}$ is defined as $\alpha$$_{\lambda_{1}-\lambda_{2}}$ = $\frac{log(\lambda_{1}F_{\lambda_{1}})-log(\lambda_{2}F_{\lambda_{2}})}{log(\lambda_{1})-log(\lambda_{2})}$, and $\lambda$ is measured in $\mu$m and F$_{\lambda}$ in erg $\cdot$ s$^{-1}$ $\cdot$ cm$^{-2}$. 

The K-band flux comes from 2MASS \citep{Skrutskie2006}, the 12 $\mu$m flux from WISE and the 70 $\mu$m one from PACS. Some exceptions to this rule were applied when these fluxes were unavailable for certain objects. Sz95 has no flux at 70 $\mu$m, so we used PACS100 $\mu$m instead. We did the same for 2MASS J1608149. An exceptional case is object SSTc2dJ160111.6-413730, discussed later on section \ref{specialcase}, as no PACS flux was available, and therefore could not be classified.

\begin{figure}[htbp]
 \caption{Identification of transitional disks candidates applying the
 color criteria from \cite{Ribas2013} to the Lupus \textit{Herschel} data.}
  \centering
    \includegraphics[width=0.5\textwidth]{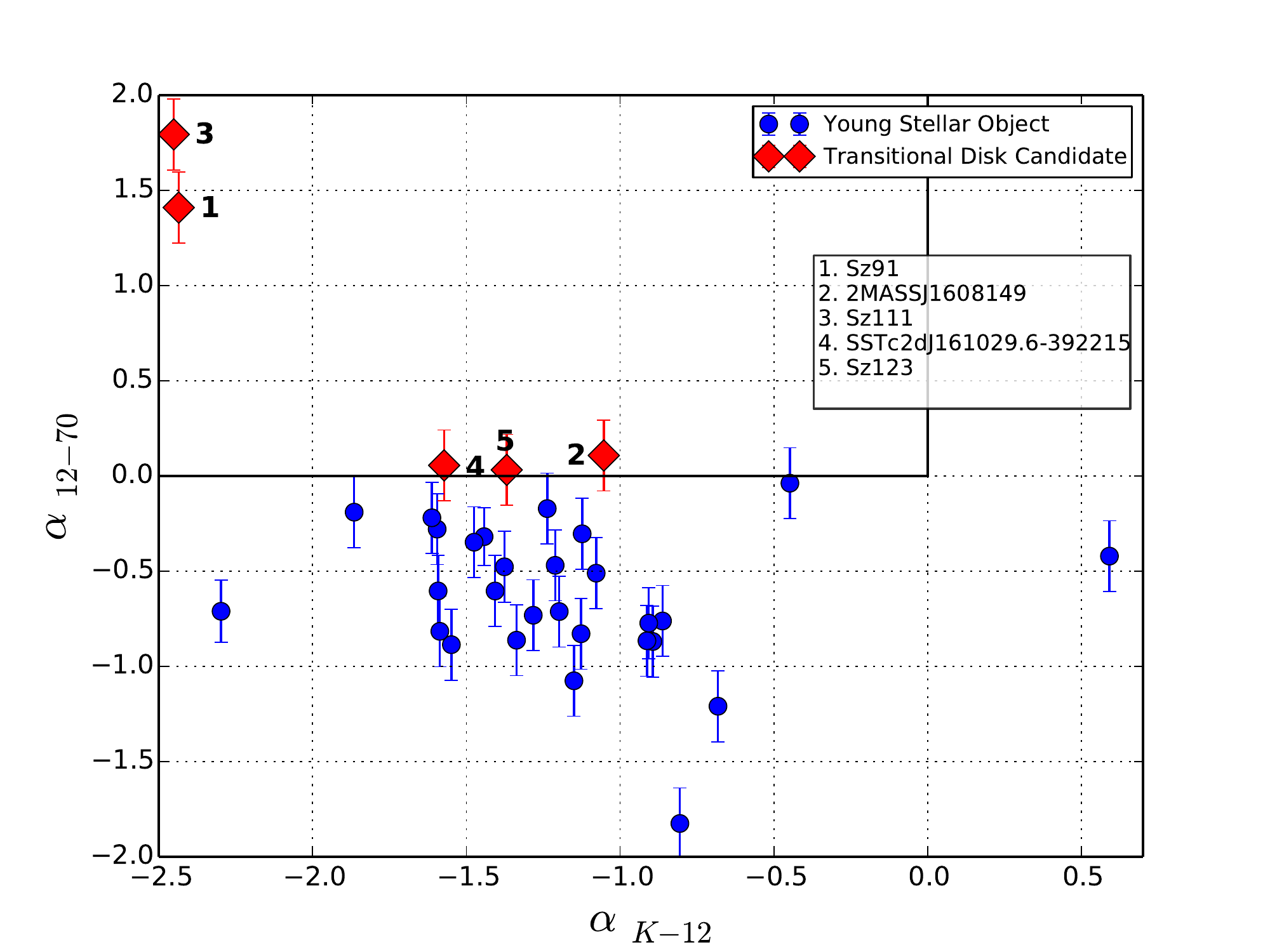}
  \label{fig:slope_plot}
\end{figure}

With this procedure we detected five objects that satisfy the transitional disk candidate criteria: SSTc2dJ161029.6-392215, Sz91, Sz111, Sz123 and 2MASS J1608149. In Figure \ref{fig:slope_plot} we represent the objects that fulfilled our selection criteria. Objects Sz91 and Sz111 were previously known transitional disks from \textit{Spitzer} data \citep{Merin2008}, and they occupy a different region in the diagram than the others. They present little to no excess at 12 $\mu$m and a steep rise between 12 and 70 $\mu$m in their SEDs (see Figure \ref{fig:seds_TD}). This could be indicative of a large cavity in both cases, mostly empty of small dust grains. In the case of Sz91, a resolved image of this inner hole has already been obtained in the sub-millimeter with SMA \citep{Tsukagoshi2014}, being one of the largest imaged so far ($\sim$65 AU) in a T Tauri disk. Sz111 is likely to have a similarly large and empty cavity.

The rest of the objects present excesses at intermediate wavelengths, as well as fulfilling our color criteria (see Figure \ref{fig:seds_TD} for a representation of their SEDs). Objects 2MASS J1608149 is a special cases, as explained at subsection \ref{specialcase2}.

In Figure \ref{fig:slope_plot} three additional objects present error bars falling in the transitional disk detection criteria: SSTc2dJ160703.9-391112, RXJ1608.6-3922 and Sz108B. Their SEDs present far infrared excesses, but mid-infrared ones as well. As they do not strictly fulfill our selection criteria, they are not to be considered transitional disk candidates in this work, yet they deserve attention.

Analysis on the remaining sample of T Tauris in the Lupus regions is outside the scope of this work, although some special cases are discussed in some detail in \S~\ref{objects}.

\subsection{Spectral Energy Distributions}
\label{seds}
To construct the SEDs we used ancillary data from ground-based optical, near-infrared, \textit{Spitzer} and \textit{WISE} data \citep{Mortier2011, Comeron2008, Allen2007, Krautter1997}. These photometric data provide continuous wavelength coverage between 0.4 and 500 $\mu$m. Figures \ref{fig:seds_TD}, \ref{fig:seds1} and \ref{fig:seds2} show the SEDs of the transitional disk candidates and the rest of the detected objects, respectively. We used the interstellar extinction law of \cite{Weingartner2001}. The \textit{A$_{V}$} values were extracted from \cite{Mortier2011} and \cite{Merin2008}, except in 7 sources, identified with an asterisk in Table \ref{tab:table2}, for which we derived them from the observed optical and near-infrared photometry and their spectral types.

We inspected all the images in 2MASS band J of the entire sample of detected objects. All of them were confirmed as point sources except those marked in Table \ref{tab:stellarparameters} with a \textit{j}, where minor extended emission cannot be discarded.

\subsection{Individual objects of interest}
\label{objects}
\subsubsection{Sz91 and Sz111}
Sz91 and Sz111 had been previously identified as potential transitional disks based on \textit{Spitzer} data in \cite{Merin2008} and have large H$\alpha$ equivalent widths of 95.9 and 145.2 $\AA$ \citep{Hughes1994}, which indicate youth and active gas accretion to the stars.

The optical images of Sz91 reveal a known companion at approximately 8$\arcsec$ \citep{Ghez1997}, which corresponds to 1200 to 1600 AU at probably the distance of the star. However, the centroids of all the \textit{Herschel}-detected sources match the coordinates of the primary source, Sz91 (see Figure \ref{fig:images}), so we assume no contribution from the companion to the measured fluxes. 

\subsubsection{SSTc2dJ161029.6-392215 and Sz123}
Objects SSTc2dJ161029.6-392215 and Sz123 present SEDs characteristic of transitional disks but with narrow gaps, instead of the wide ones found at Sz91 and Sz111. They also present small excesses at near-infrared wavelengths (3-10 $\mu$m), that could signal the presence of an inner optically thick disk.

\subsubsection{2MASS J1608149}
\label{specialcase2}
Since object 2MASS J1608149 was not detected on the PACS70 band, we then used the PACS100 one to classify it. The target then fulfilled the transitional disk detection criteria. As its SED shows \ref{fig:seds_TD}, the \textit{Herschel} far-infrared excess is not as significant as to definitely classify the object as a transitional disk.

A further analysis including modeling could reveal the true nature of this object. As for this work, we maintain object 2MASS J1608149 as a transitional disk candidate.

\subsubsection{Sz68}
Figure \ref{fig:images} shows an extended feature at all wavelengths to the North of the Sz68 object. This is an already known Herbig-Haro object driven by Sz68 \citep{Moreno-Corral1995}. Interestingly, its emission dominates at wavelengths longer than 100 $\mu$m, which probably contaminates the SED of this object with unrelated flux to the star itself. A similar effect was discovered around the star T54 in Chamaeleon I \citep{Matra2012}, erroneously classified before as a transitional disk due to nearby extended emission \cite[see also][]{Lestrade2012}.

\subsubsection{SSTc2dJ160111.6-413730}
\label{specialcase}
SSTc2dJ160111.6-413730 has a very peculiar SED (see Fig. \ref{fig:seds2}) with no excess at any wavelengths between optical and 160 $\mu$m but substantial potential excesses at 250 to 500 $\mu$m associated to a clear point source in the images (see Fig. \ref{fig:images5}). The low 70 $\mu$m upper limit is nevertheless consistent with the sensitivity limit of the observations.

Previous claims of such extreme type of objects have been made in the context of the DUNES \textit{Herschel} Key Programme, who named these objects "cold disks'' to make reference to their extreme nature \citep{Krivov2013}. Other authors have pointed towards background galaxies as the possible explanation \citep{Gaspar2014}. It did not fulfill our classification criteria as transitional disk candidate.

\subsubsection{Sz65 and Sz66}
\label{Sz65Sz66}
Objects Sz65 and Sz66 form a binary system, with a separation of 6.4'' \citep{Lommen2010}, unresolved by \textit{Herschel} (see Figure \ref{fig:images}). We inspected their SEDs to select which one of them was the main contributor. We deduce the primary source to be Sz65 by studying the contour diagram of the image. The contour lines were centered at the position of source Sz65. Their MIPS24 and MIPS70 fluxes support this association. Then we set the \textit{Herschel} fluxes of Sz66 as upper limits in Table \ref{tab:table2} and in Figure \ref{fig:seds1}.

\section{Discussion}
\label{discussion}
\subsection{Detection statistics}
From an initial sample of 217 known Class II sources with spectroscopic membership confirmation in the Lupus clouds, 123 objects fell in at least one of the fields observed by the \textit{Herschel} Gould Belt Survey. In particular, 92 and 98 were in the PACS and SPIRE Lupus III maps, respectively; 12 and 15 were found in Lupus I PACS and SPIRE mosaics; and 10 objects were observed in Lupus IV with both instruments. From these sources, we detected 35 objects, 32 in PACS70, 2 in PACS100 and 1 in SPIRE. We then discarded one, 2MASSJ1608537, as we could not unequivocally assign its \textit{Herschel} photometry to the target.

Compared with a similar study in Chamaeleon \citep{Ribas2013}, here we report a \textit{Herschel} detection rate of $\sim$ 28\% of the previously known objects in the region, comparable with the percentage in that work ($\sim$30\%). The small different distances to the regions ($\sim$ 150 pc for Chamaeleon vs $\sim$ 200 pc to Lupus III) might account for the small difference in the detection rate. Further analysis of the detection statistics of known and new sources in these regions with the HGBS data will be presented by Benedittini et al. (in prep.).

\subsection{Incidence of transitional disks in Lupus}

We report the detection of five transitional disk candidates in the Lupus molecular clouds based on \textit{Herschel} and previous data of all spectroscopically confirmed Class II members of the association. Two transitional disks were known from previous works (Sz91 and Sz111) but the other three are new to the literature.

The Spitzer study of the Lupus clouds found a global disk fraction of 50 - 60\% \citep{Ribas2014}, compatible with a relatively young age for the region of 1 - 2 Myrs \citep{Comeron2008}. However, the total sample in \cite{Merin2008} did include objects without spectroscopic membership confirmation so it could have certain level of contamination from background galaxies and/or old dusty AGB stars \cite[see][for a spectroscopic survey of Spitzer selected candidate members in Lupus]{Mortier2011}. This work however uses an input target list where all objects have spectroscopic membership confirmation and is therefore not affected by that. 

The observed fraction of transitional disks in Lupus based on our \textit{Herschel} detected sample is $\sim$15\%, which is comparable to the fractions measured in other such young and nearby regions \citep{Espaillat2014}. This corrects a surprisingly low incidence of this type of objects from the Spitzer sample reported in \cite{Merin2008}, where only 2 objects were identified as transitional, from a photometric sample of 139 Class II and III sources. This demonstrates how \textit{Herschel} efficiently complements mid-infrared surveys for the specific study of transitional disks.

\subsection{Brighter PACS-70 fluxes in transitional disks}

\begin{figure*}[HT]    
 \centering
 \caption{SEDs of the 5 transitional disk candidates identified with \textit{Herschel}. Blue dots represent the dereddened ancillary data from the literature from previous studies. The red diamonds are the clear detections with \textit{Herschel}, and the red triangles the upper limits wherever no clear detection in the maps was found. The gray solid line is the median SED from the Lupus sample of the 29 non transitional disks in the region, and the gray shaded area the first and fourth quartiles. See Table \ref{tab:mediansed} for more information. The dashed lines are the photospheric NextGen models from \cite{Allard2012}.}
  \includegraphics[width=0.4\textwidth,page=1]{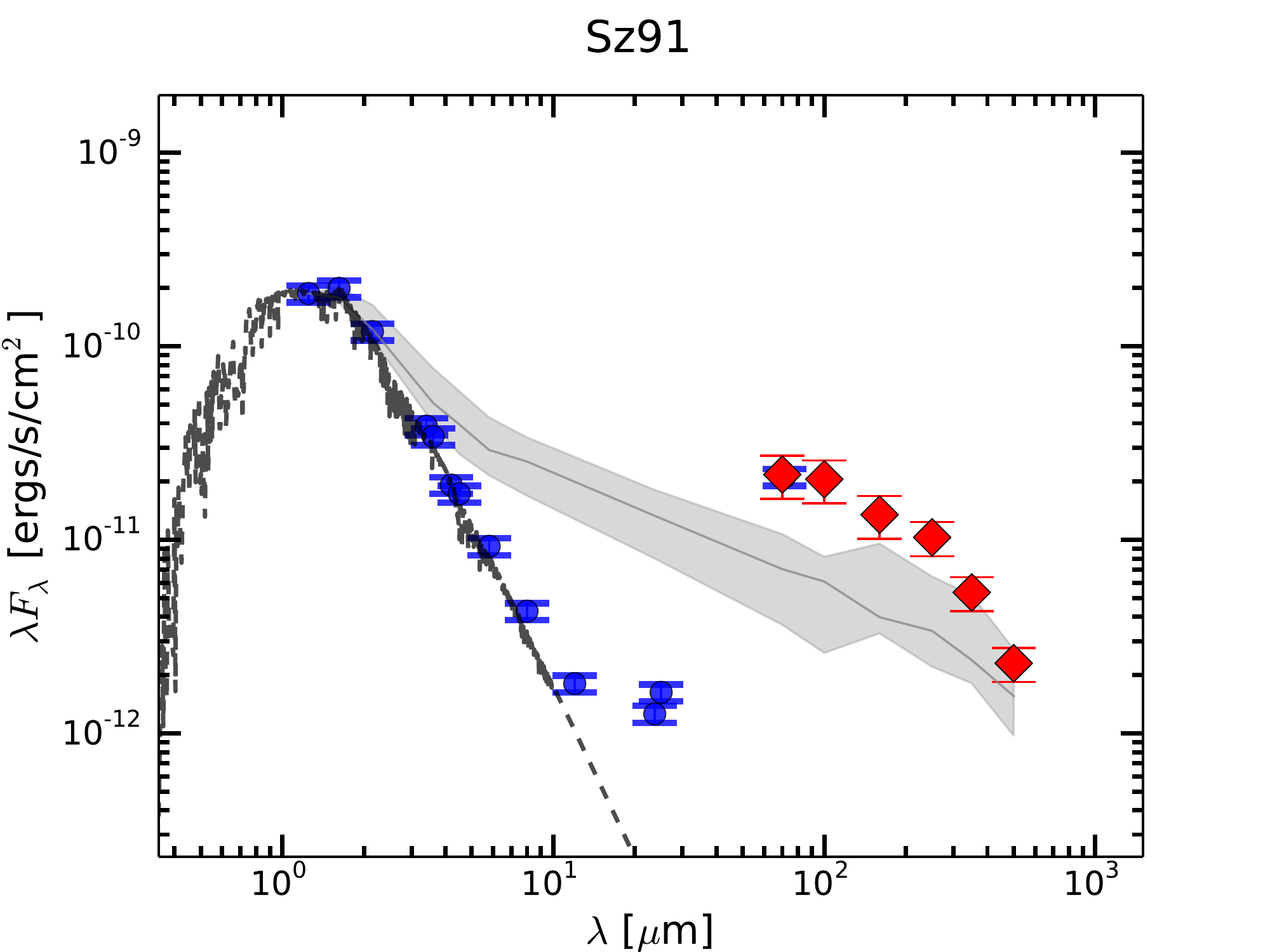}
  \includegraphics[width=0.4\textwidth,page=2]{Seds_TD.pdf}
\includegraphics[width=0.4\textwidth,page=3]{Seds_TD.pdf}
\includegraphics[width=0.4\textwidth,page=4]{Seds_TD.pdf}
  \includegraphics[width=0.4\textwidth,page=5]{Seds_TD.pdf}
   \label{fig:seds_TD}
\end{figure*}

\begin{figure*}[HT]    
 \centering
 \caption{SEDs of the 29 YSOs with \textit{Herschel} detections that do not fulfill our transitional disk selection criteria}
  \includegraphics[width=0.3\textwidth,page=1]{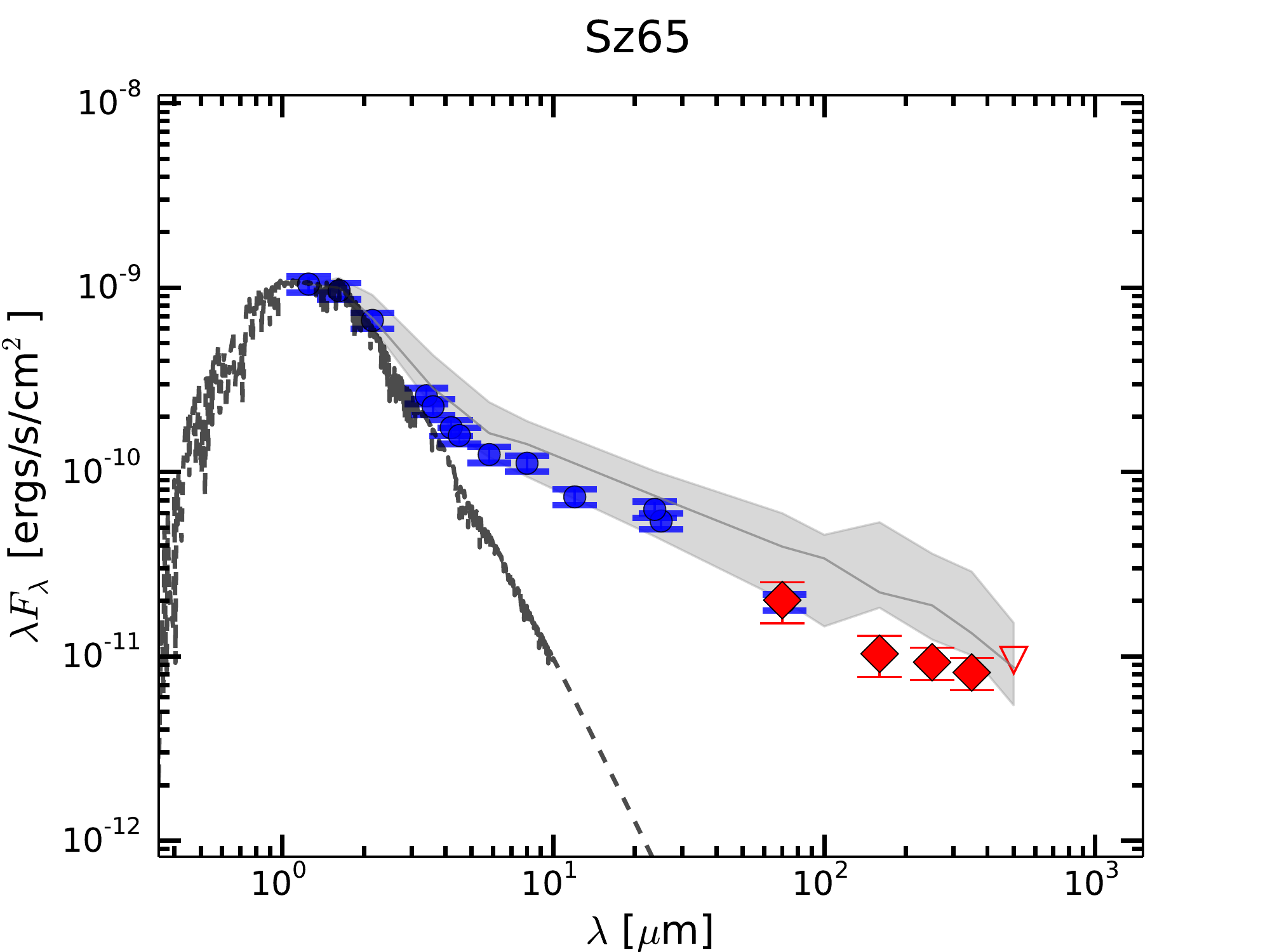}
  \includegraphics[width=0.3\textwidth,page=2]{Seds_rest.pdf}
\includegraphics[width=0.3\textwidth,page=3]{Seds_rest.pdf}
\includegraphics[width=0.3\textwidth,page=4]{Seds_rest.pdf}
  \includegraphics[width=0.3\textwidth,page=5]{Seds_rest.pdf}
  \includegraphics[width=0.3\textwidth,page=6]{Seds_rest.pdf}
\includegraphics[width=0.3\textwidth,page=7]{Seds_rest.pdf}
\includegraphics[width=0.3\textwidth,page=8]{Seds_rest.pdf}
  \includegraphics[width=0.3\textwidth,page=9]{Seds_rest.pdf}
  \includegraphics[width=0.3\textwidth,page=10]{Seds_rest.pdf}
\includegraphics[width=0.3\textwidth,page=11]{Seds_rest.pdf}
\includegraphics[width=0.3\textwidth,page=12]{Seds_rest.pdf}
\includegraphics[width=0.3\textwidth,page=13]{Seds_rest.pdf}
\includegraphics[width=0.3\textwidth,page=14]{Seds_rest.pdf}
\includegraphics[width=0.3\textwidth,page=15]{Seds_rest.pdf}
   \label{fig:seds1}
\end{figure*}

\begin{figure*}[HT]   
 \centering
 \caption{Fig. \ref{fig:seds1} continued.}
\includegraphics[width=0.3\textwidth,page=16]{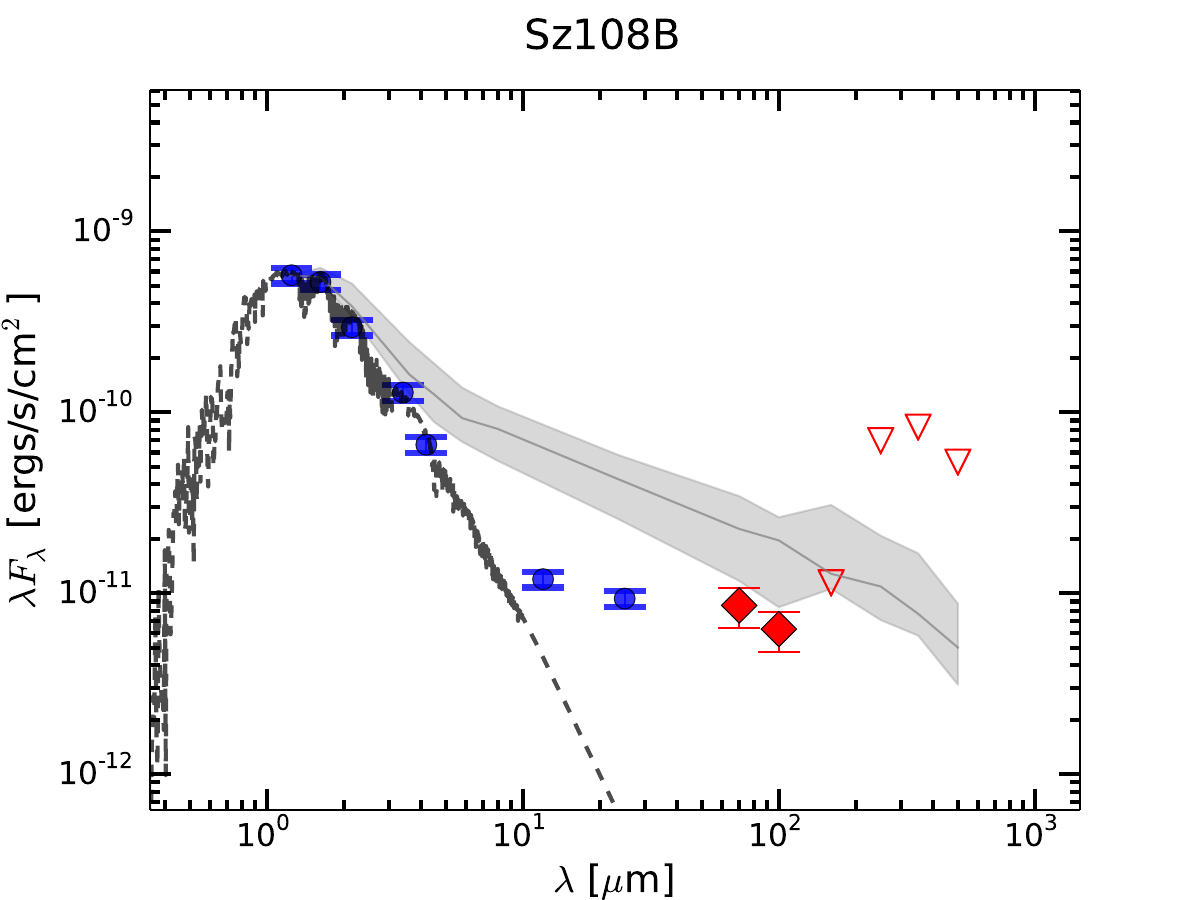}
\includegraphics[width=0.3\textwidth,page=17]{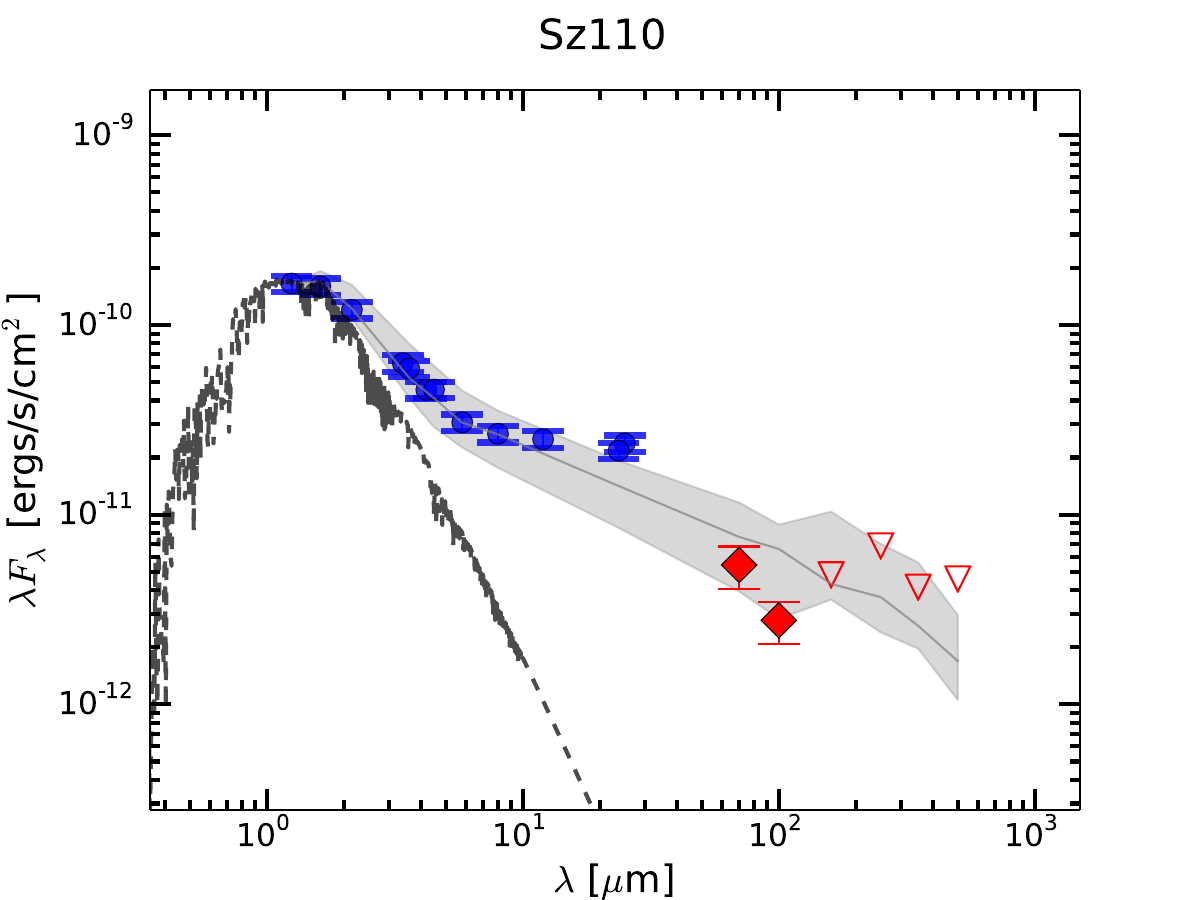}
\includegraphics[width=0.3\textwidth,page=18]{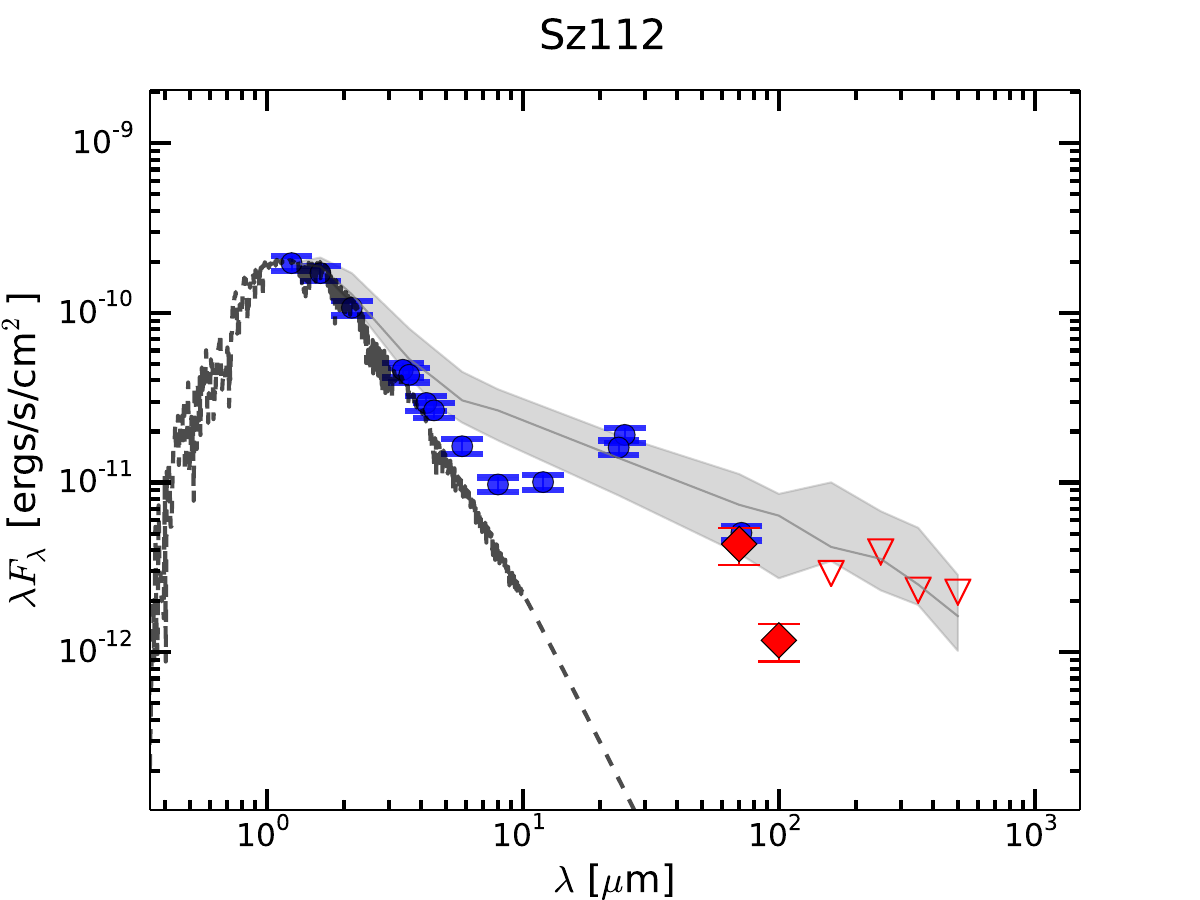}
  \includegraphics[width=0.3\textwidth,page=19]{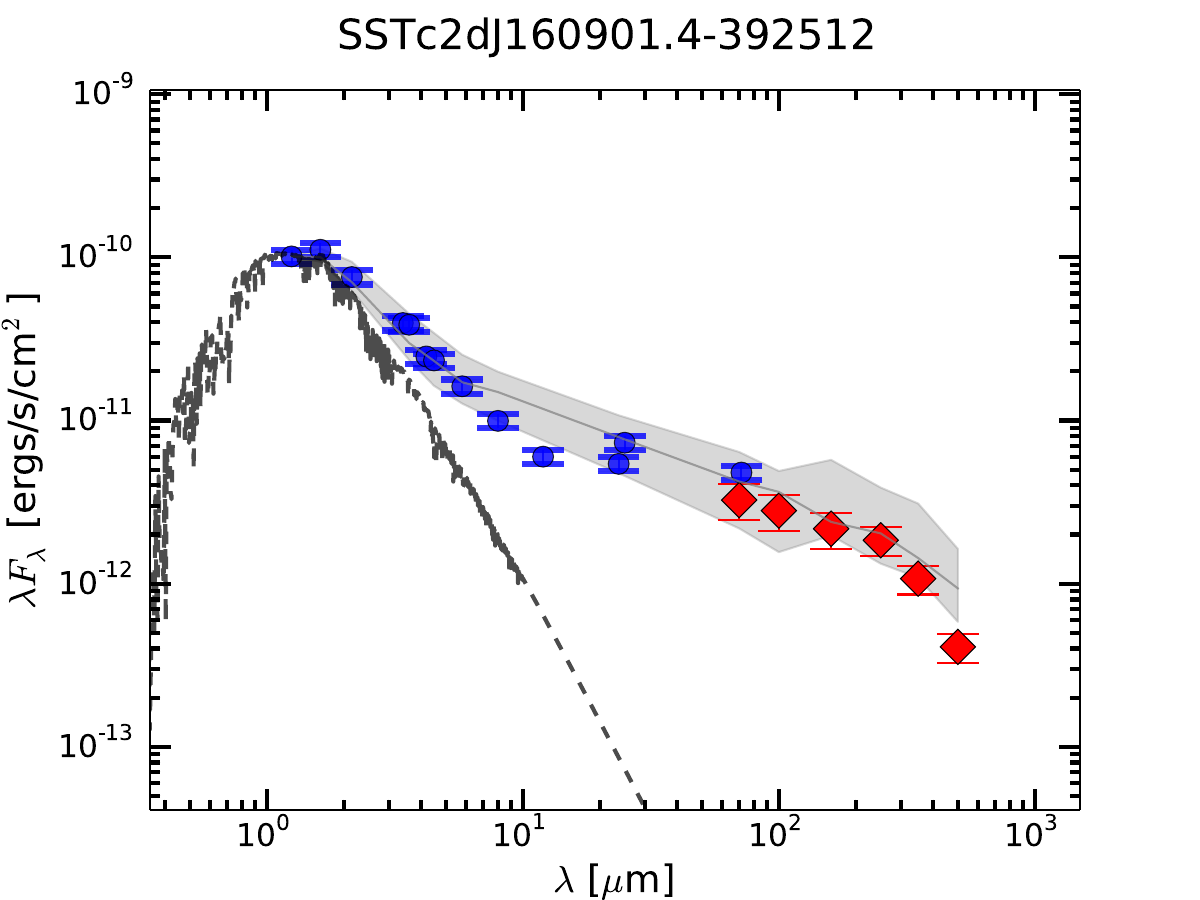}
  \includegraphics[width=0.3\textwidth,page=20]{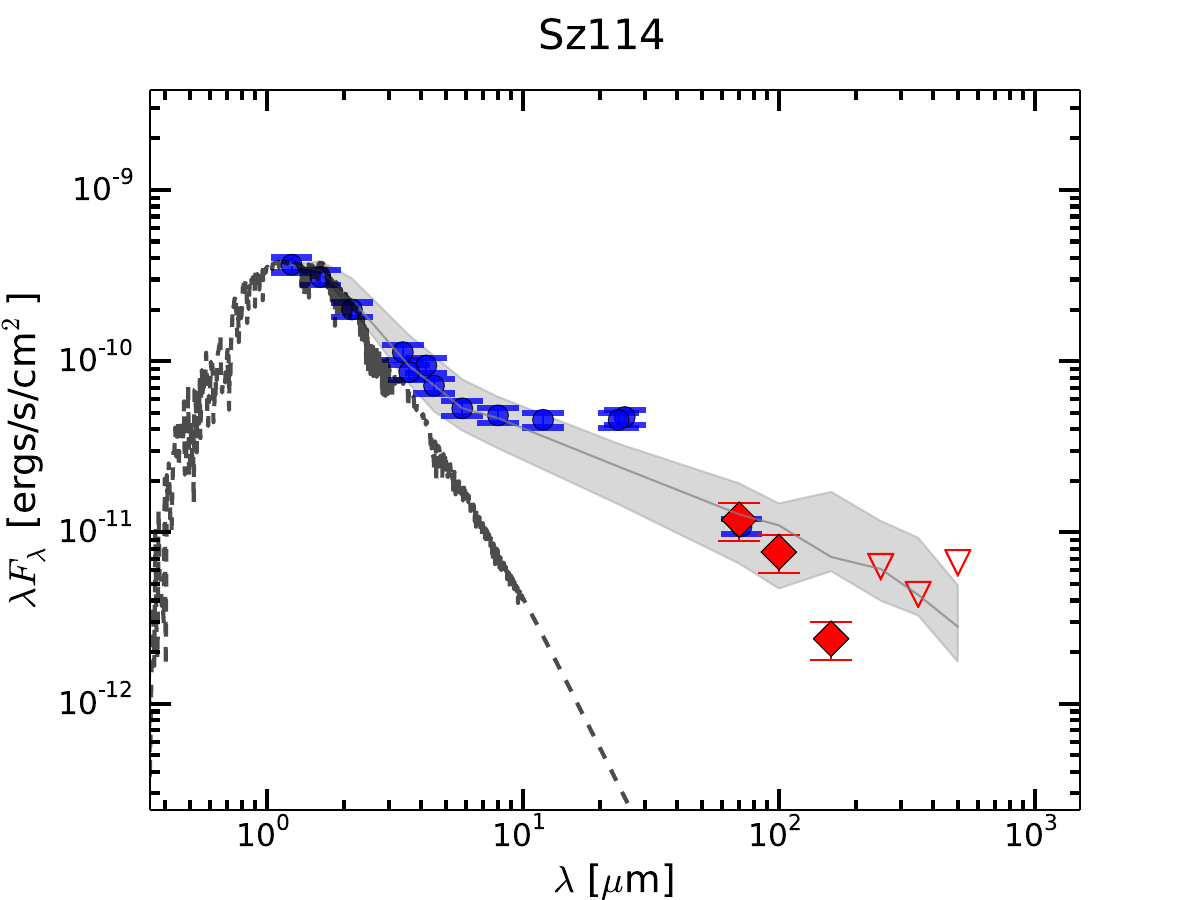}
\includegraphics[width=0.3\textwidth,page=21]{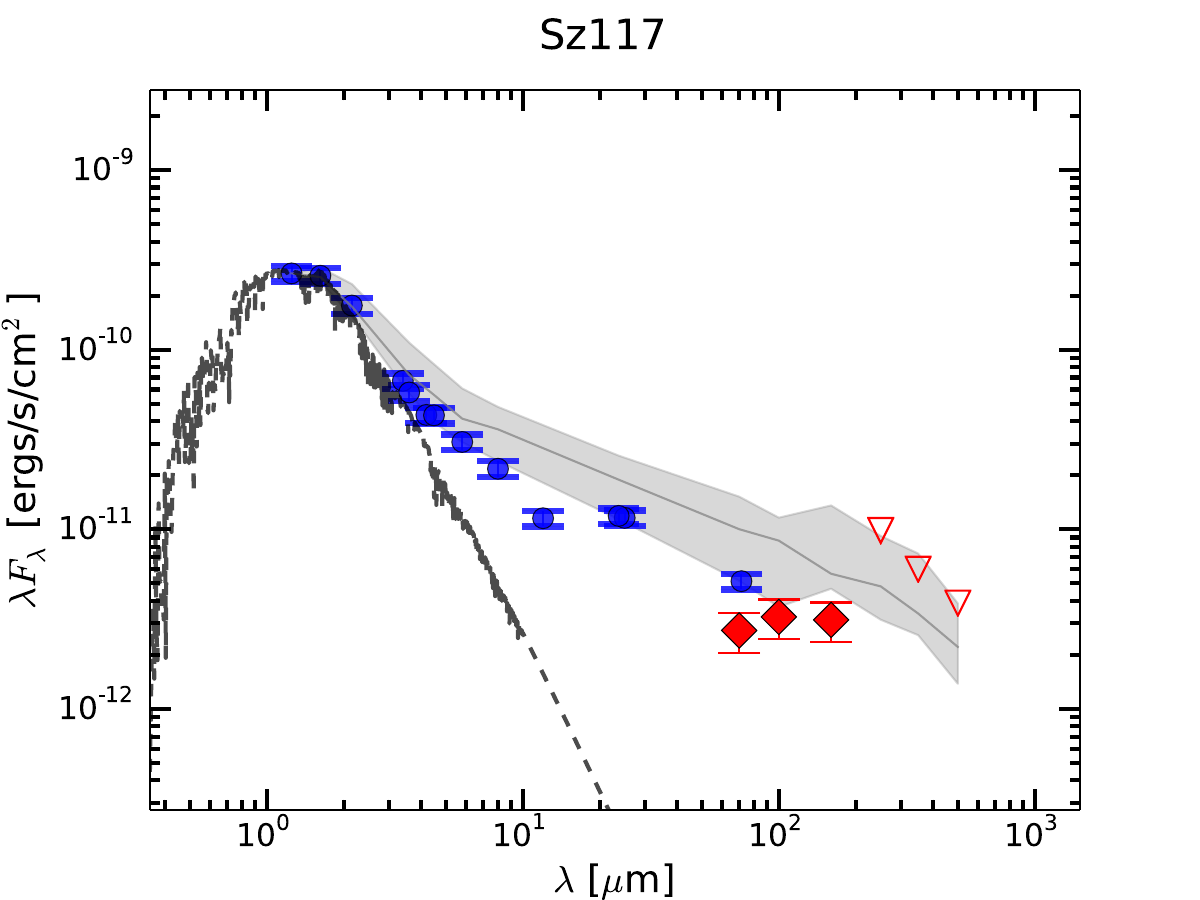}
\includegraphics[width=0.3\textwidth,page=22]{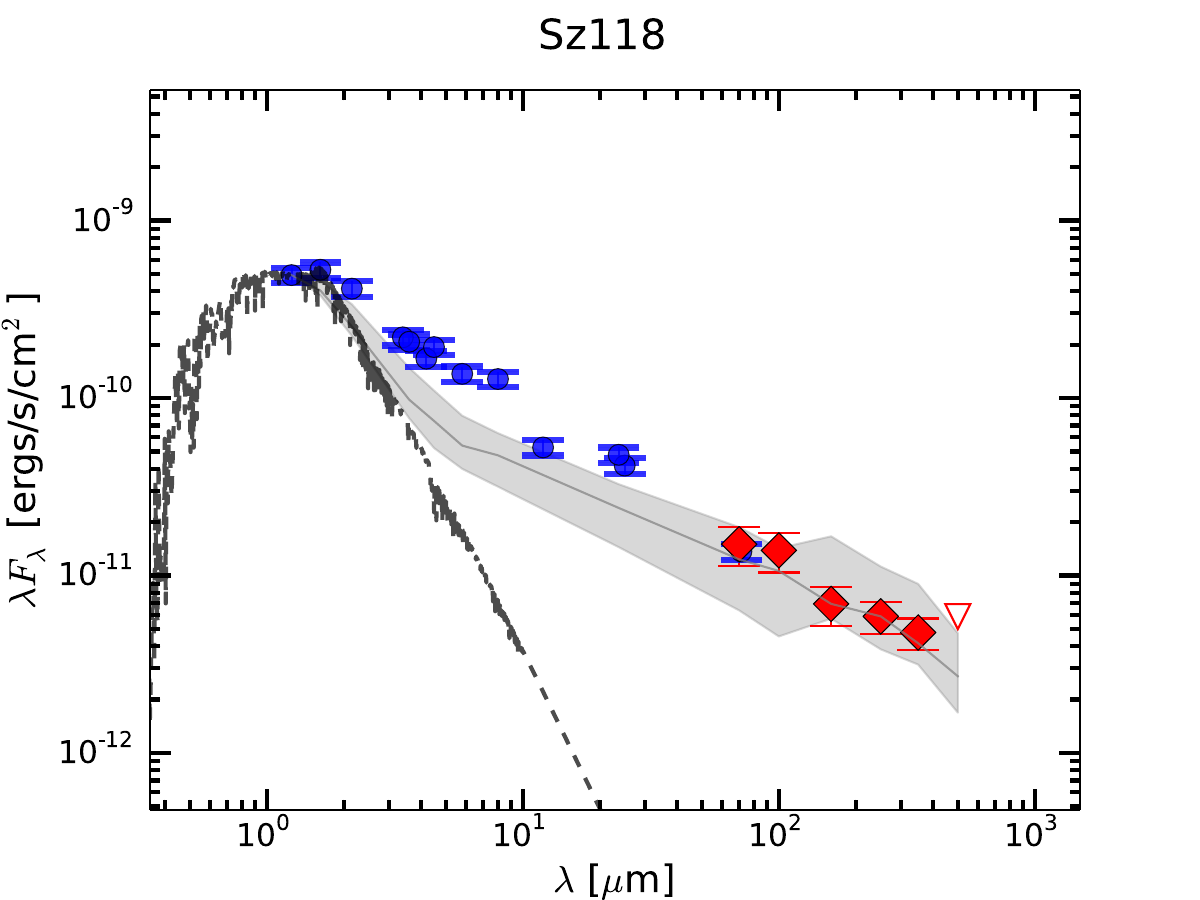}
  \includegraphics[width=0.3\textwidth,page=23]{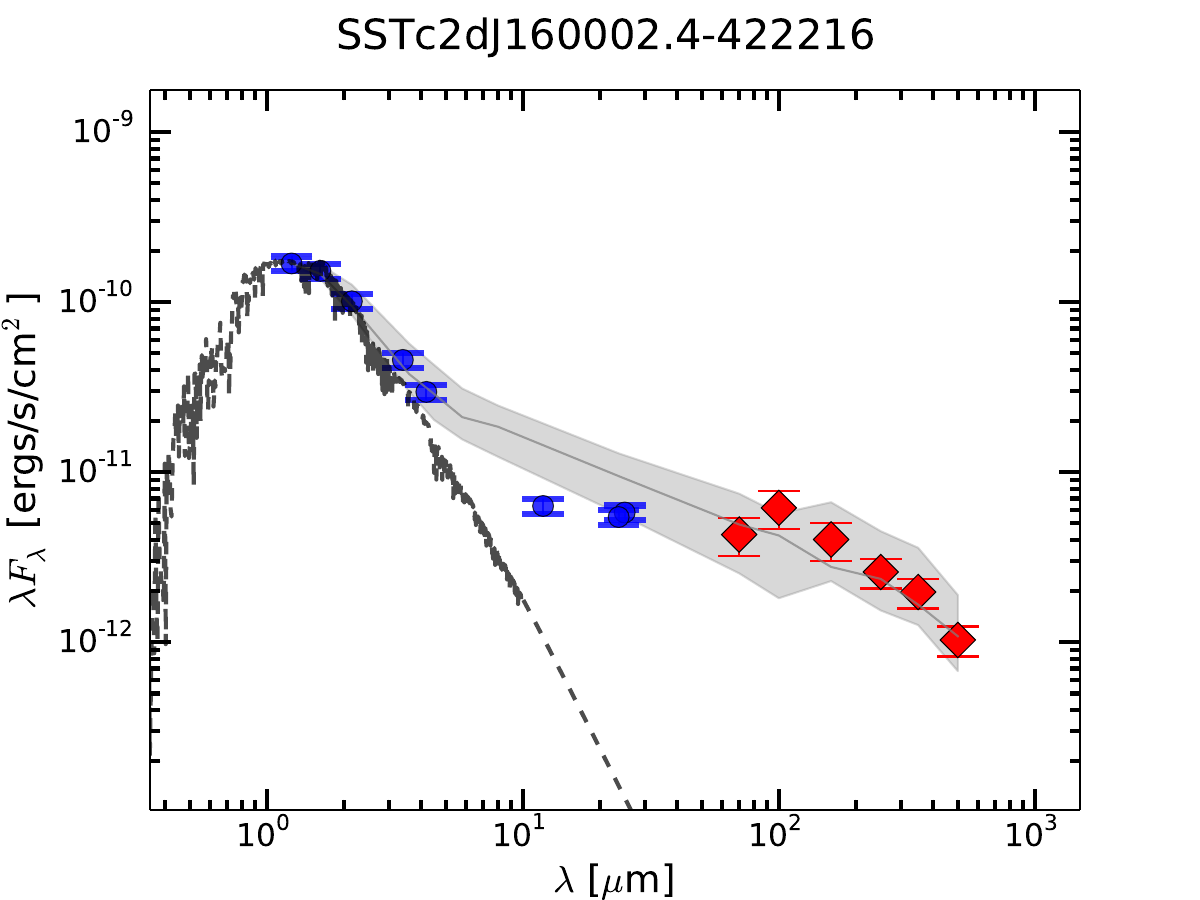}
  \includegraphics[width=0.3\textwidth,page=24]{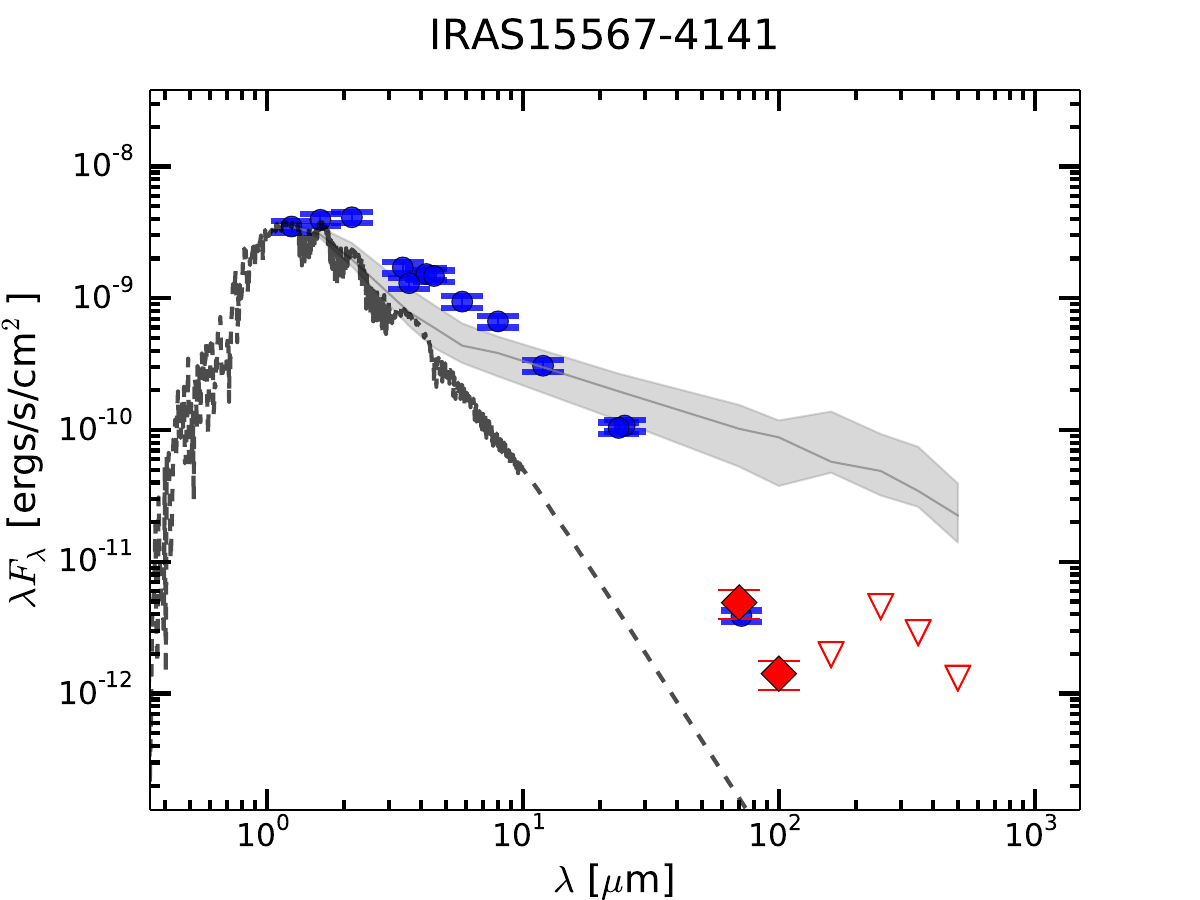}
\includegraphics[width=0.3\textwidth,page=25]{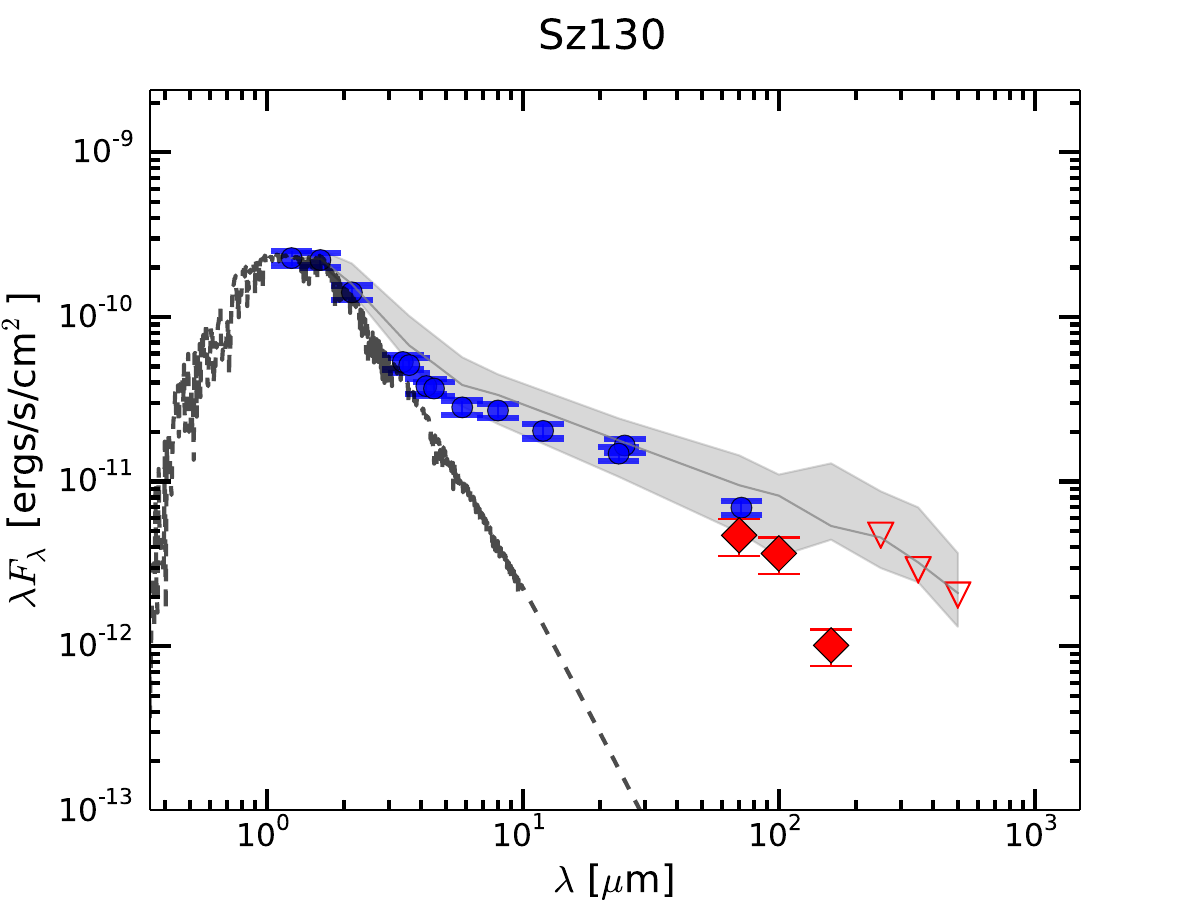}
\includegraphics[width=0.3\textwidth,page=26]{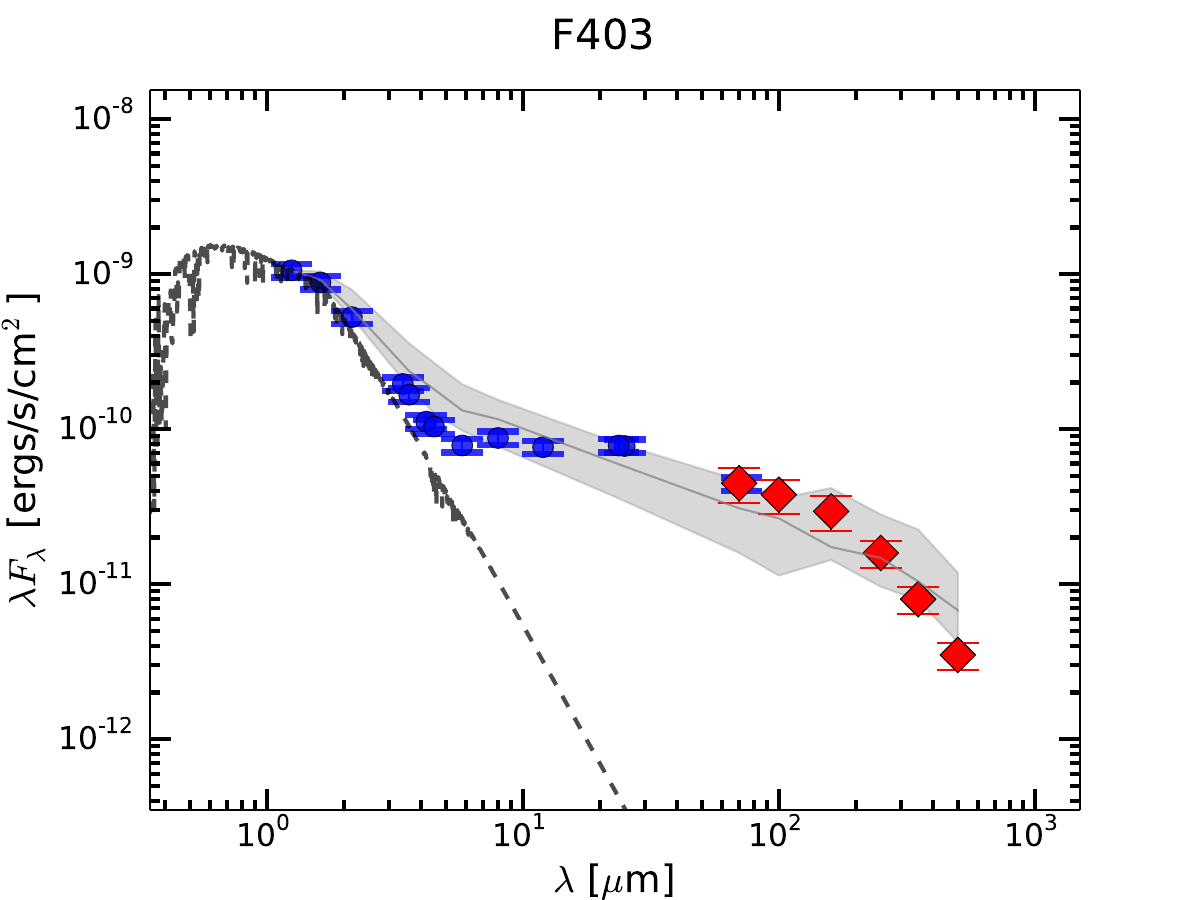}
  \includegraphics[width=0.3\textwidth,page=27]{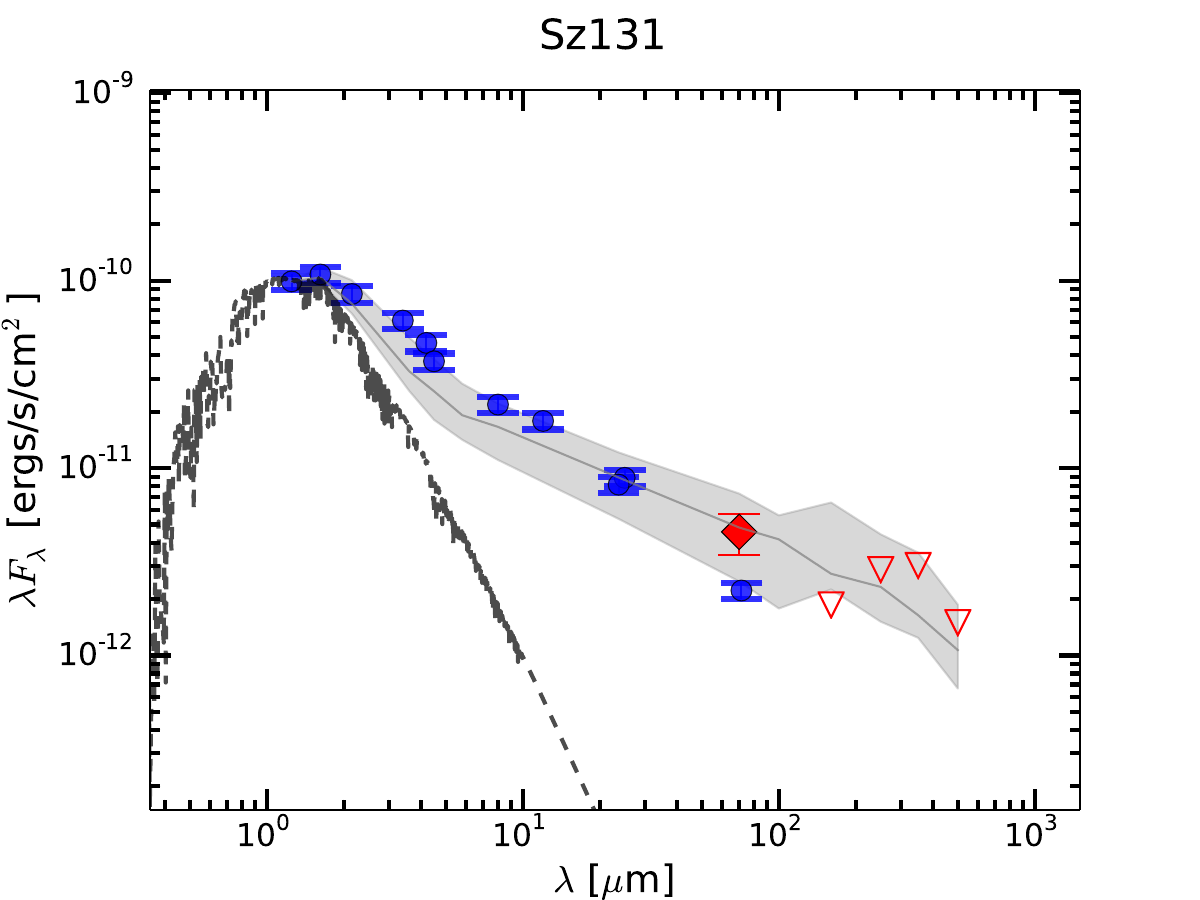}
\includegraphics[width=0.3\textwidth,page=28]{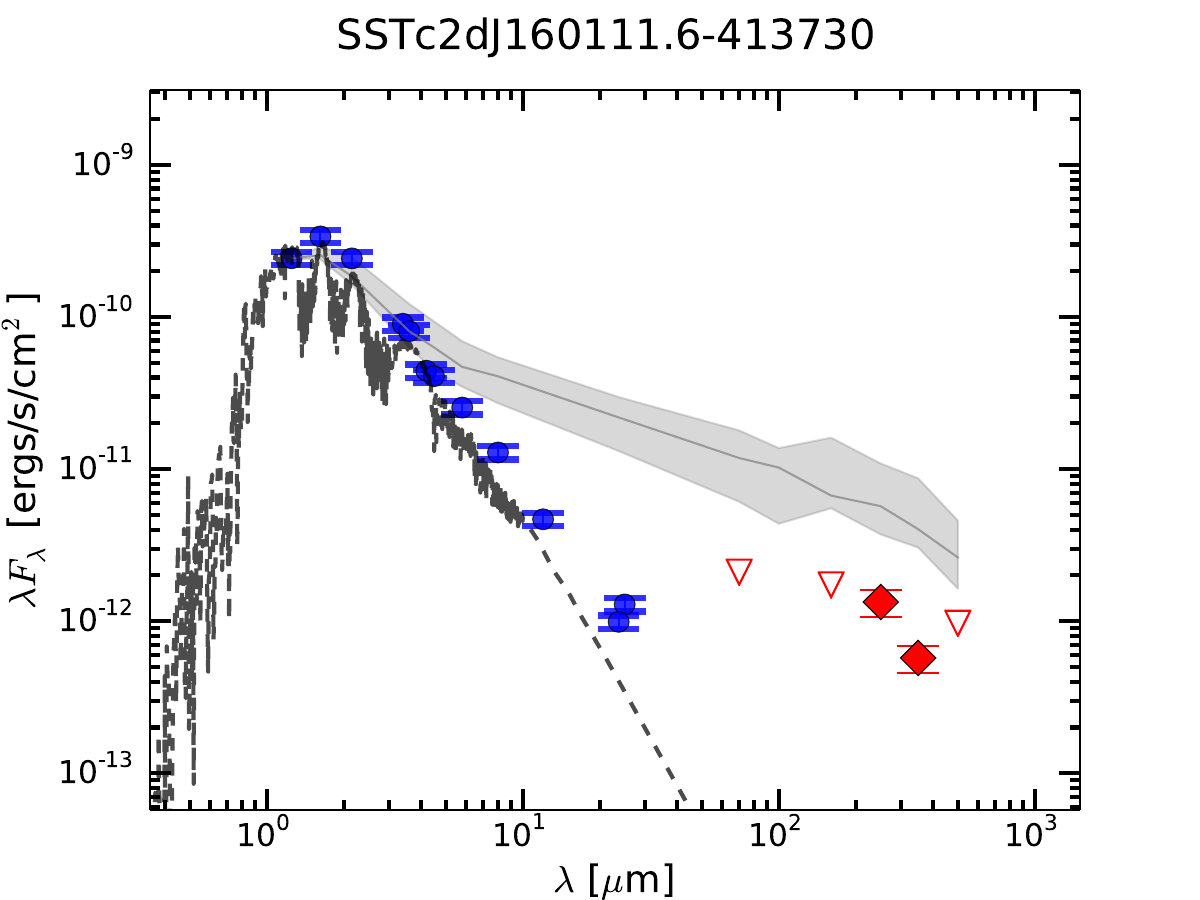}
\includegraphics[width=0.3\textwidth,page=29]{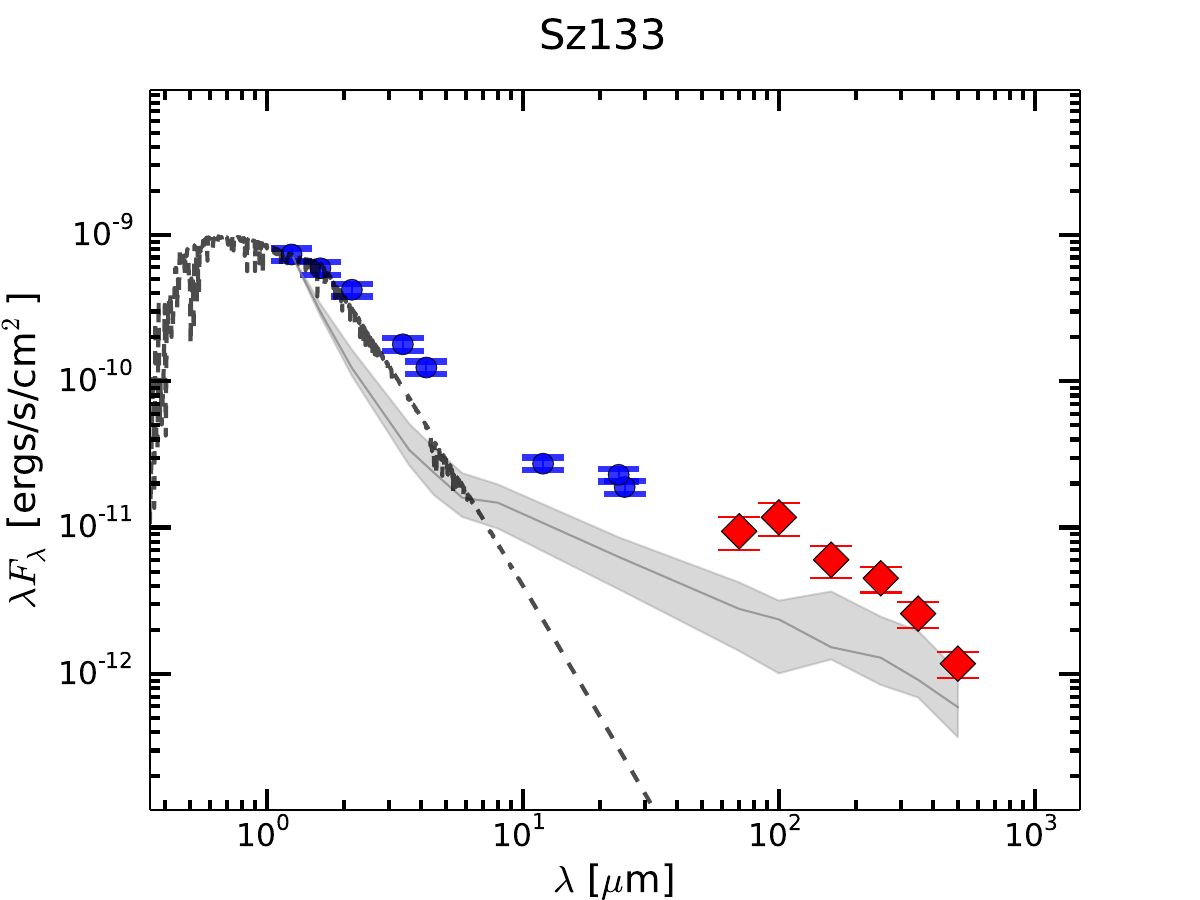}
   \label{fig:seds2}
\end{figure*}

\begin{figure*}[HT]
  \centering
 \caption{Images of each of the 34 sources with at least one point source detected by \textit{Herschel}, plus object 2MASSJ1608537, added for completeness. The images correspond to a box of 120" x 120" in size. Coordinates are given in Table \ref{tab:stellarparameters}. The color scale is defined with the RMS of the pixel values (as background level) and this number plus three times the standard deviation (as maximum level). North is up and East is left. A circle indicates the expected position of the target. The standard deviation computed around the sources is shown in the bottom right side of the images.}.
    \includegraphics[width=1\textwidth,page=1]{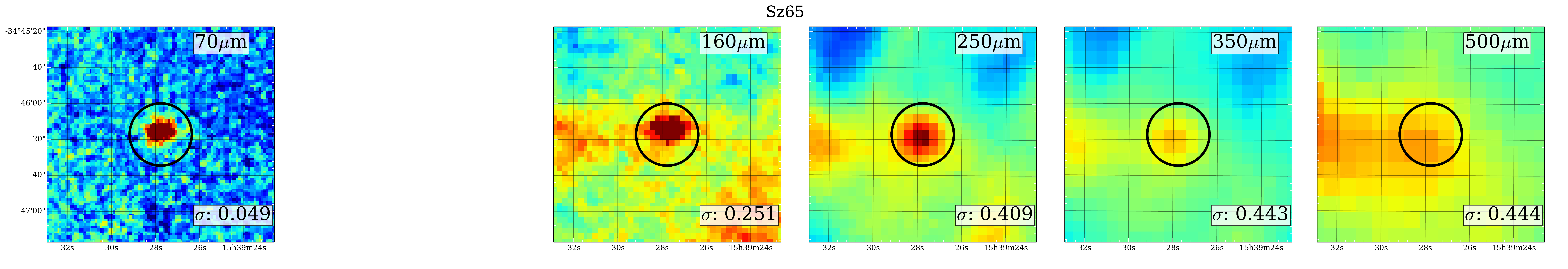}
    \includegraphics[width=1\textwidth,page=2]{Paper_Images.pdf}
    \includegraphics[width=1\textwidth,page=3]{Paper_Images.pdf}
    \includegraphics[width=1\textwidth,page=4]{Paper_Images.pdf}
    \includegraphics[width=1\textwidth,page=5]{Paper_Images.pdf}
    \includegraphics[width=1\textwidth,page=6]{Paper_Images.pdf}
    \includegraphics[width=1\textwidth,page=7]{Paper_Images.pdf}
  \label{fig:images}
\end{figure*}

\begin{figure*}[HT]
  \centering
\caption{Fig. \ref{fig:images} continued.}
    \includegraphics[width=1\textwidth,page=8]{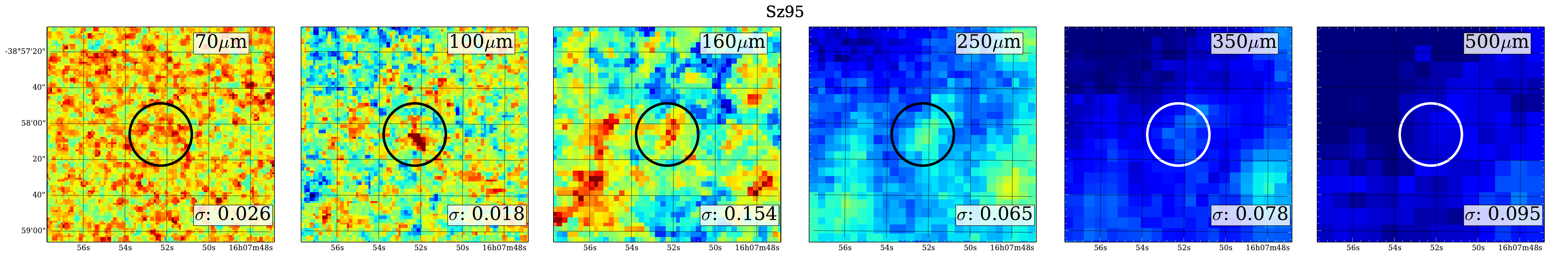}
    \includegraphics[width=1\textwidth,page=9]{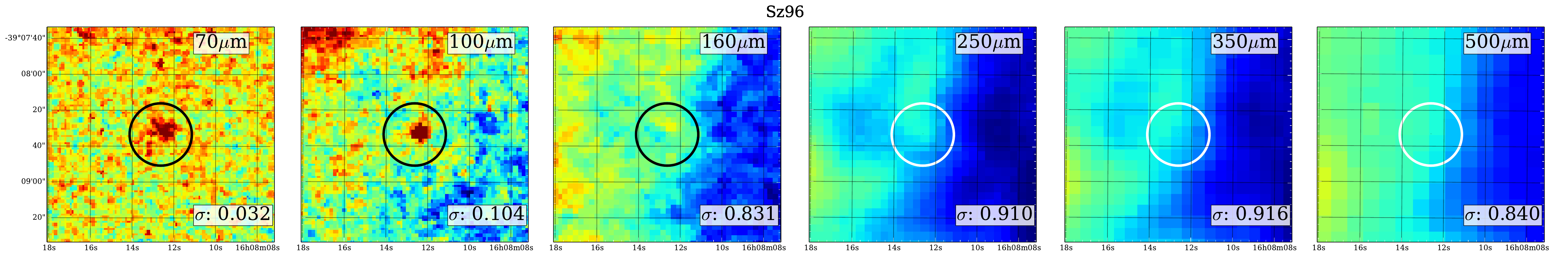}
    \includegraphics[width=1\textwidth,page=10]{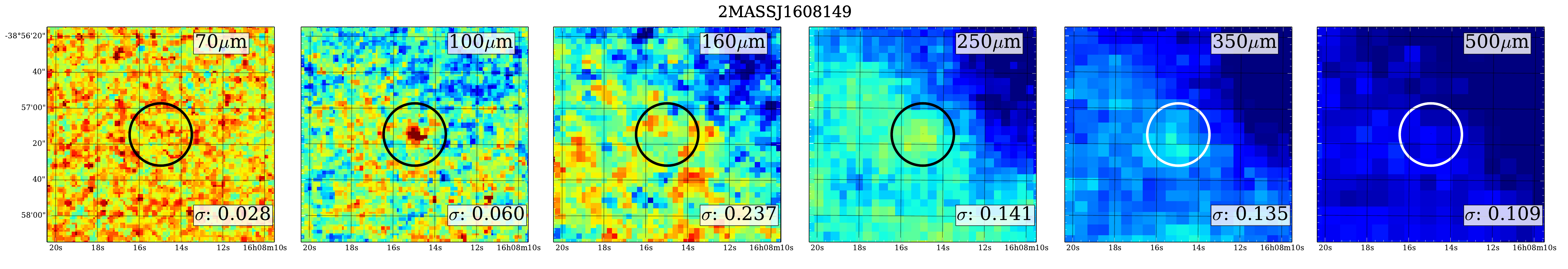}
    \includegraphics[width=1\textwidth,page=11]{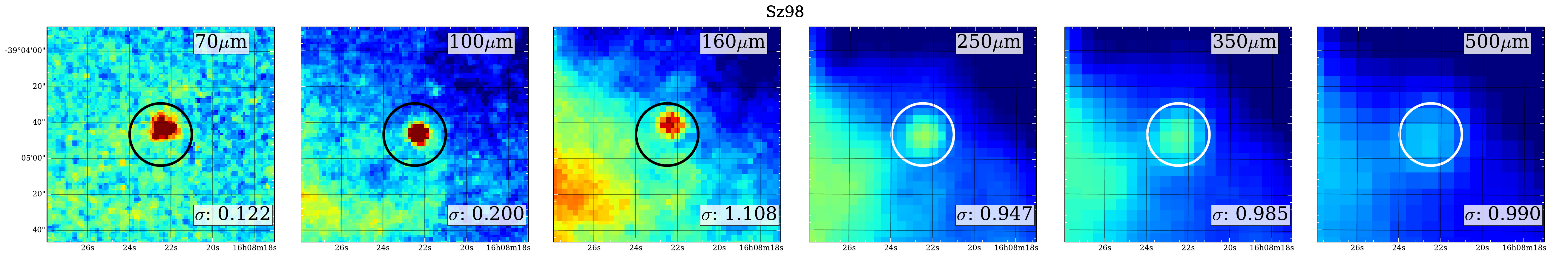}
    \includegraphics[width=1\textwidth,page=12]{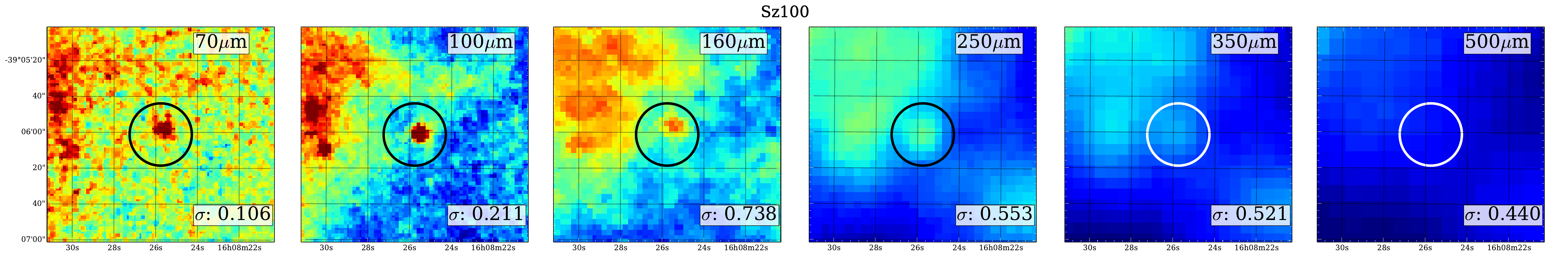}
    \includegraphics[width=1\textwidth,page=13]{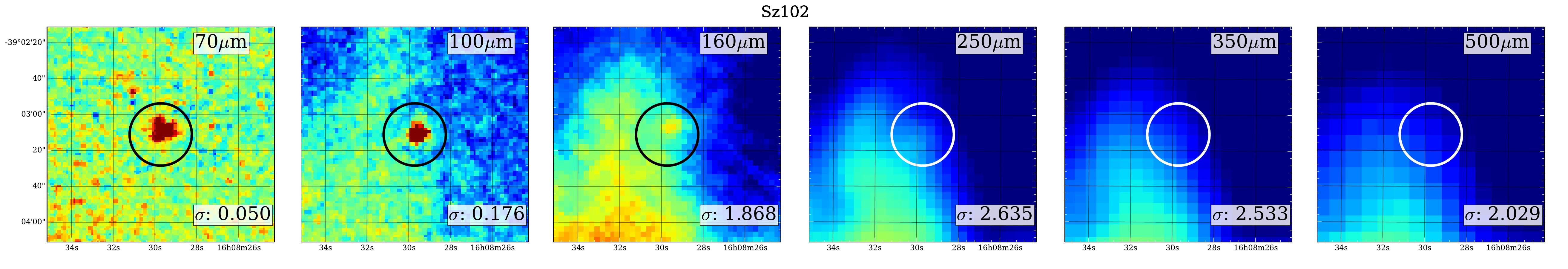}
    \includegraphics[width=1\textwidth,page=14]{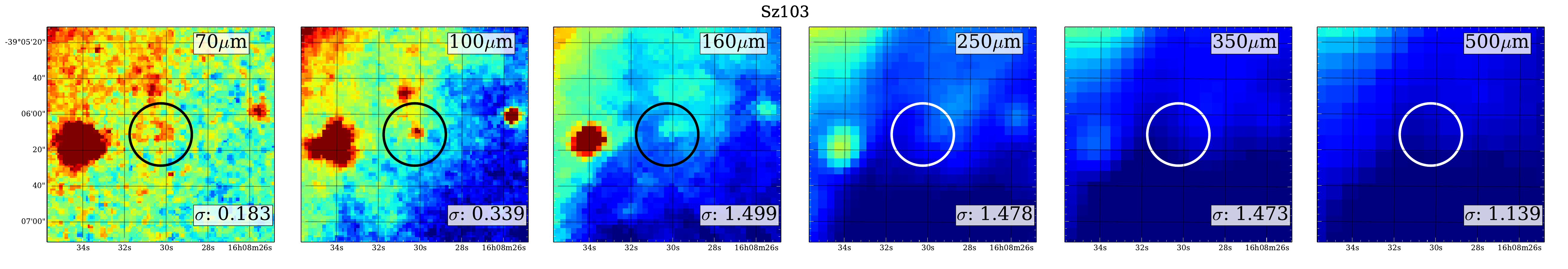}
  \label{fig:images2}
\end{figure*}

\begin{figure*}[HT]
  \centering
\caption{Fig. \ref{fig:images} continued.}
    \includegraphics[width=1\textwidth,page=15]{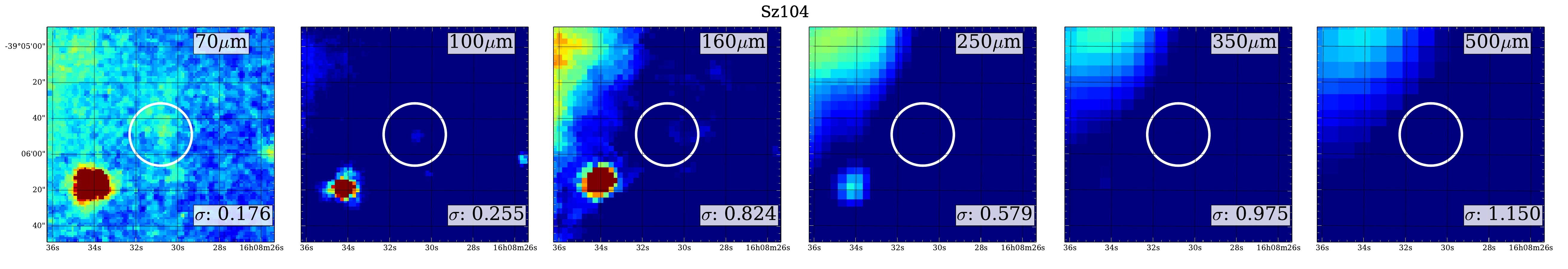}
    \includegraphics[width=1\textwidth,page=16]{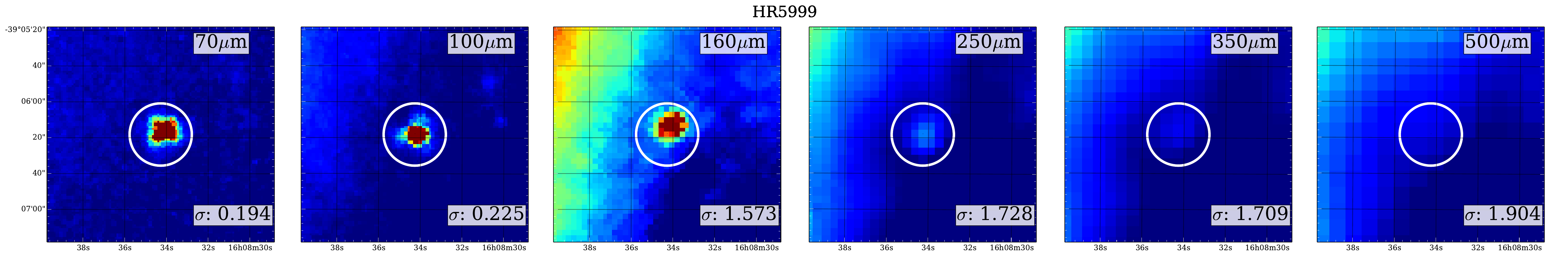}
    \includegraphics[width=1\textwidth,page=17]{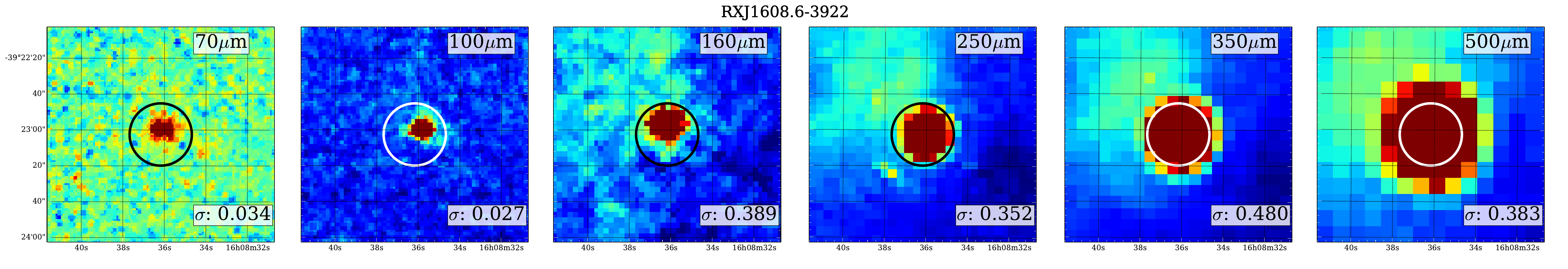}
    \includegraphics[width=1\textwidth,page=18]{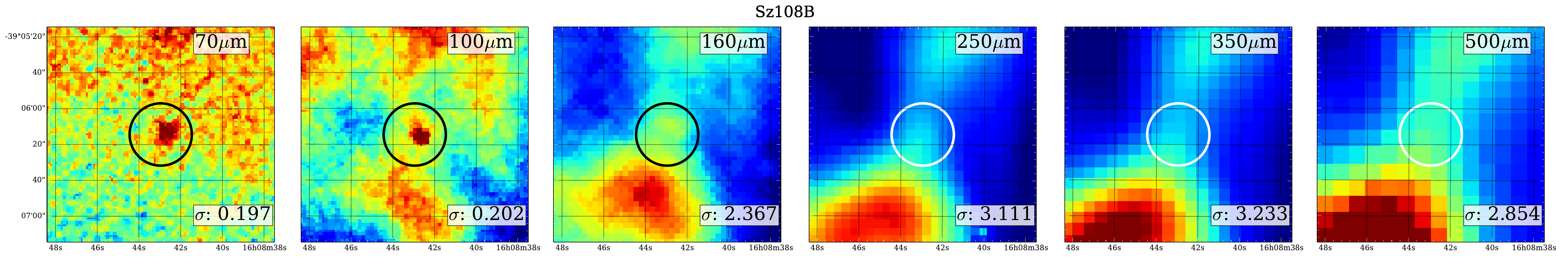}
    \includegraphics[width=1\textwidth,page=19]{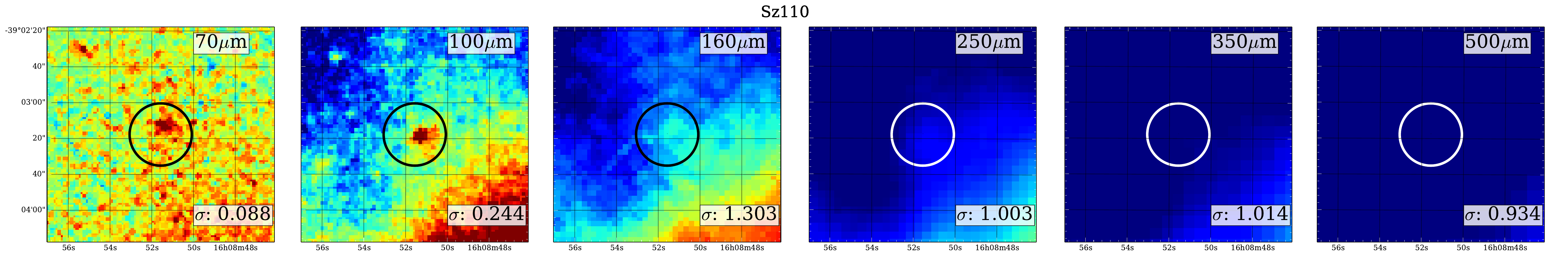}
    \includegraphics[width=1\textwidth,page=20]{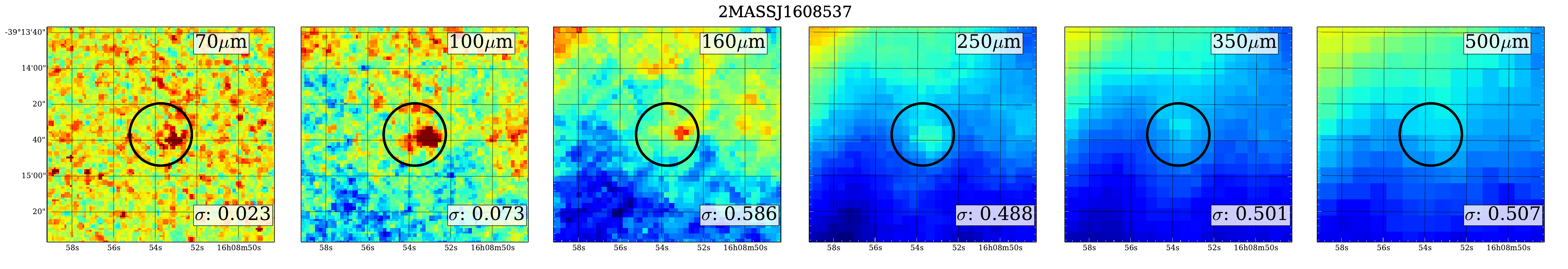}
    \includegraphics[width=1\textwidth,page=21]{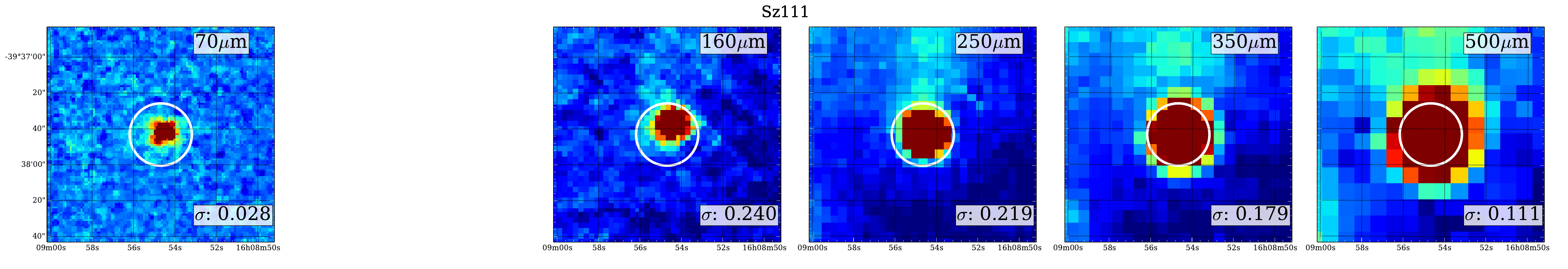}
  \label{fig:images3}
\end{figure*}

\begin{figure*}[HT]
  \centering
\caption{Fig. \ref{fig:images} continued.}
    \includegraphics[width=1\textwidth,page=22]{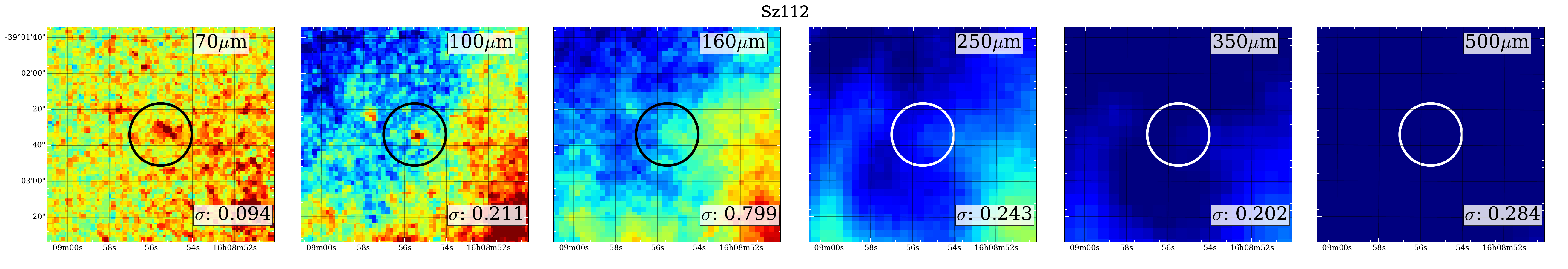}
    \includegraphics[width=1\textwidth,page=23]{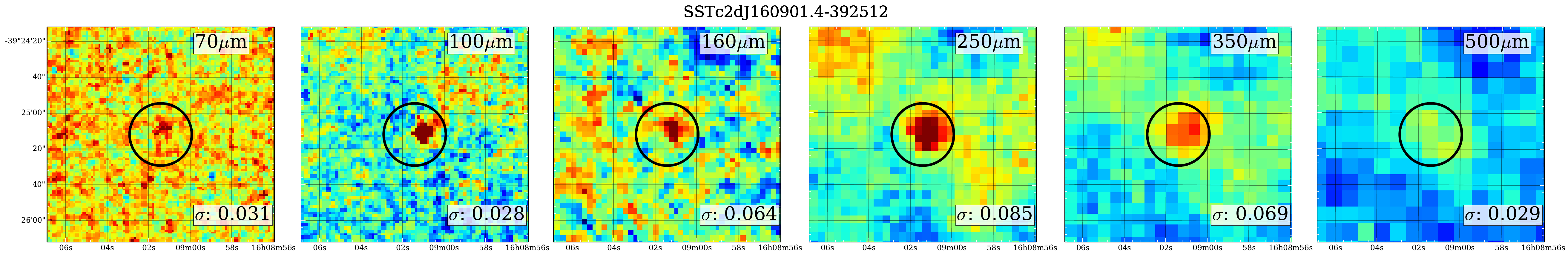}
    \includegraphics[width=1\textwidth,page=24]{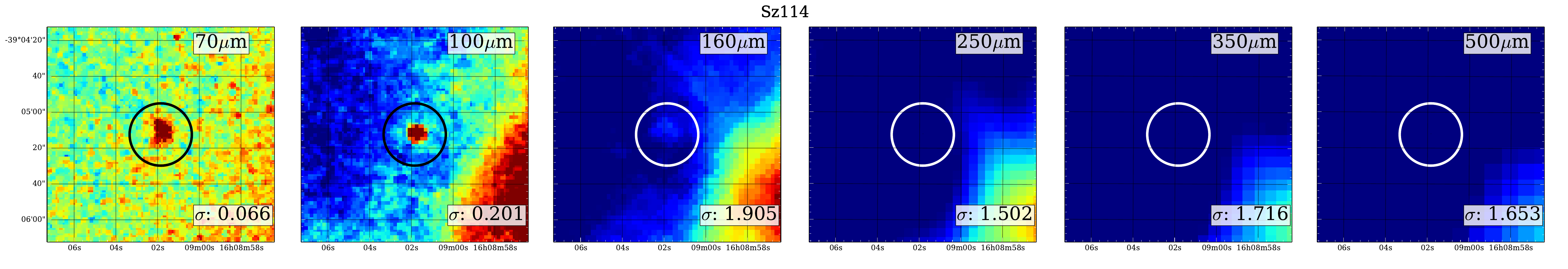}
    \includegraphics[width=1\textwidth,page=25]{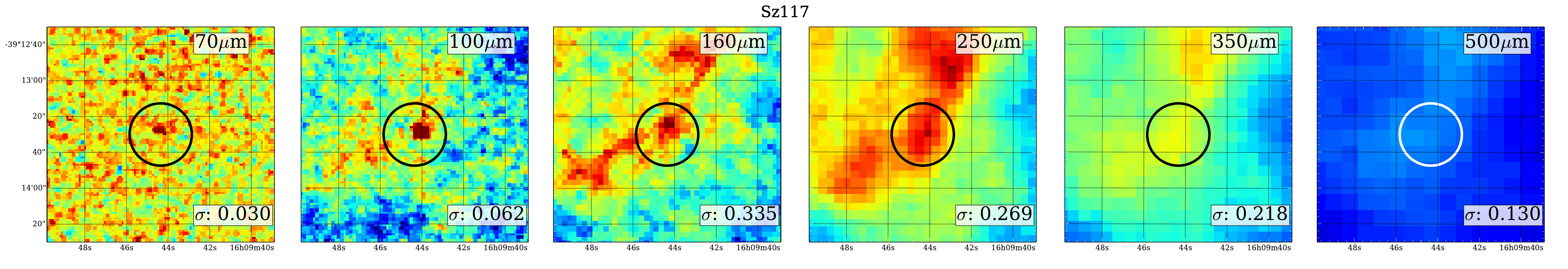}
    \includegraphics[width=1\textwidth,page=26]{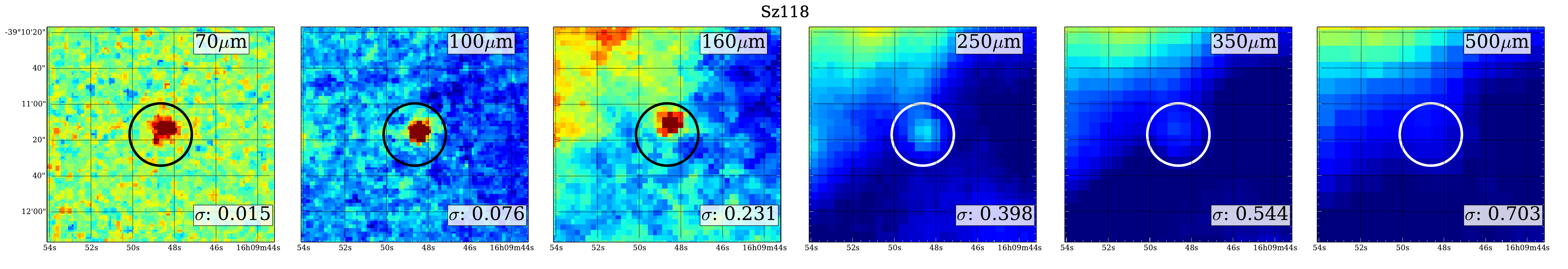}
    \includegraphics[width=1\textwidth,page=27]{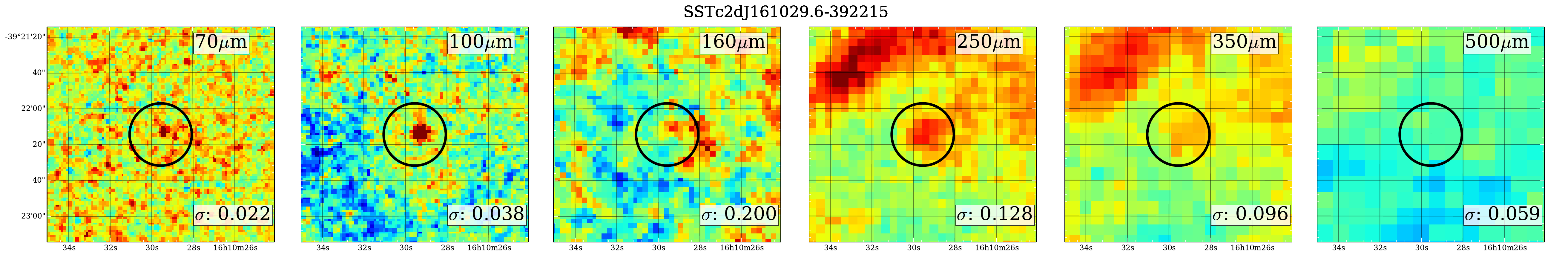}
    \includegraphics[width=1\textwidth,page=28]{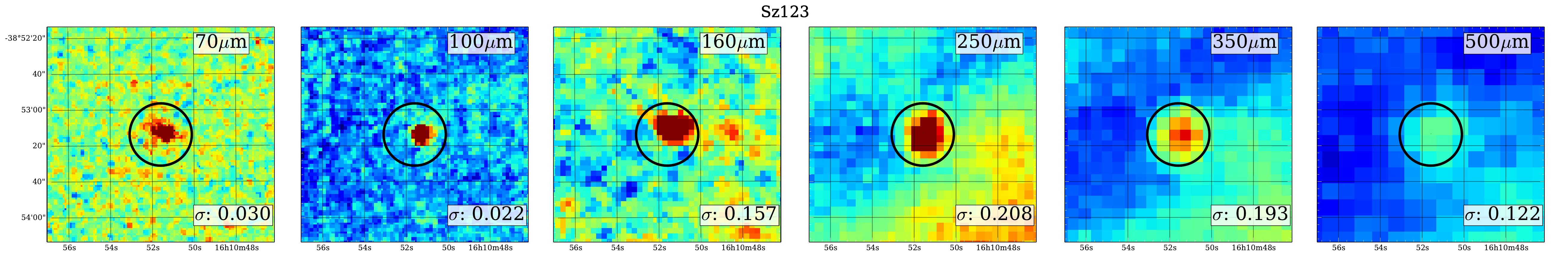}
  \label{fig:images4}
\end{figure*}

\begin{figure*}[HT]
  \centering
\caption{Fig. \ref{fig:images} continued.}
    \includegraphics[width=1\textwidth,page=29]{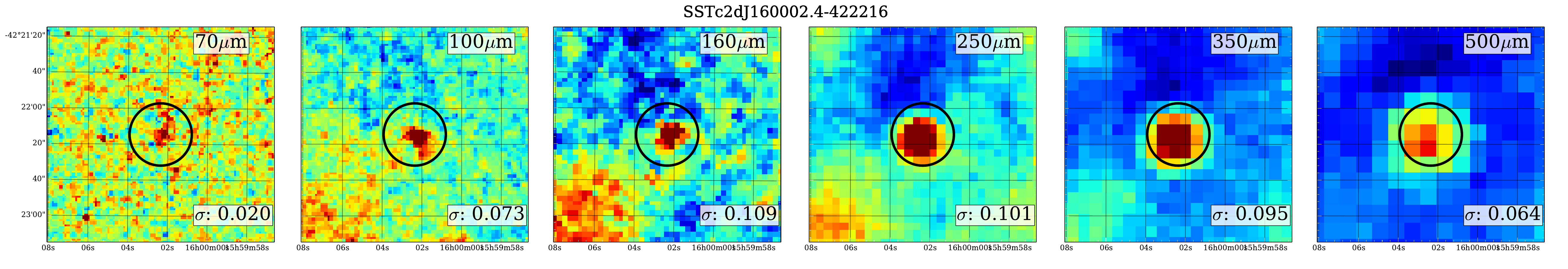}
    \includegraphics[width=1\textwidth,page=30]{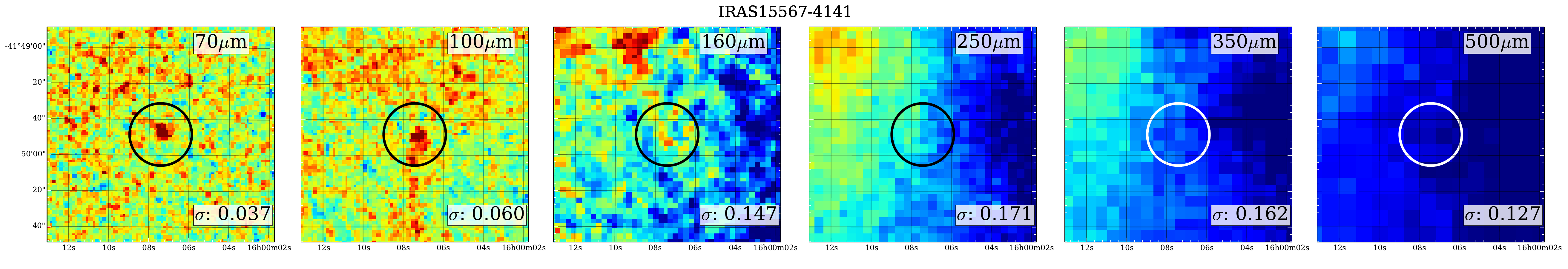}
    \includegraphics[width=1\textwidth,page=31]{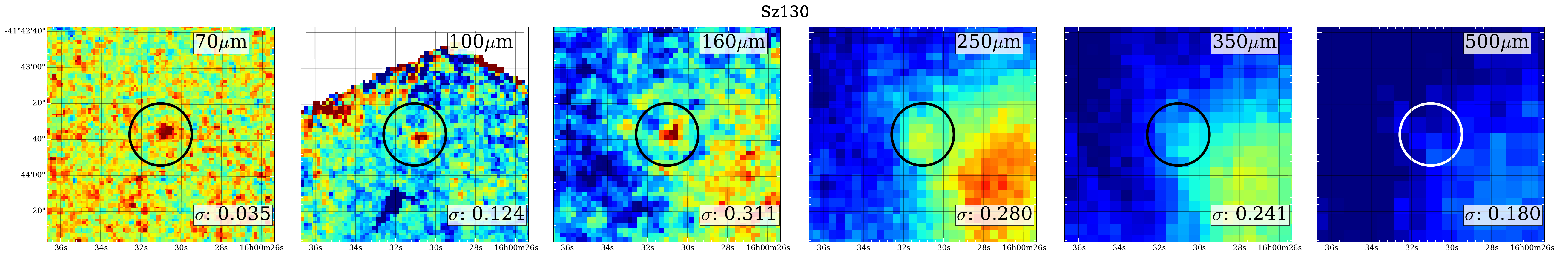}
    \includegraphics[width=1\textwidth,page=32]{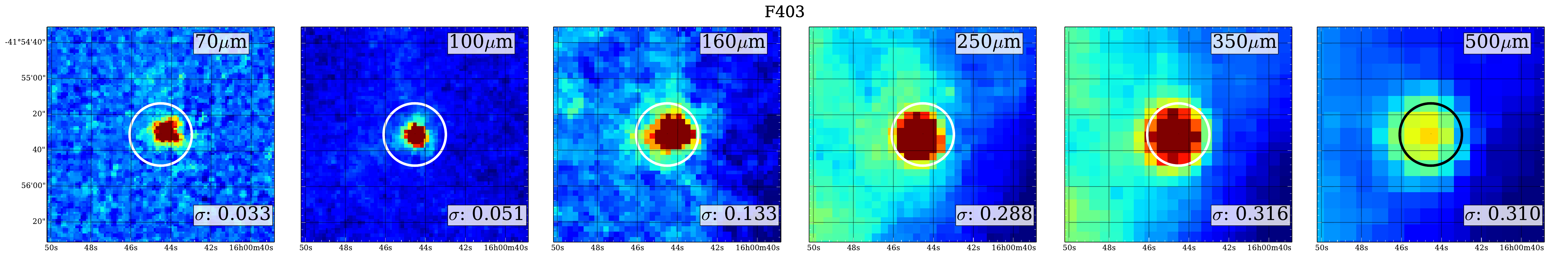}
    \includegraphics[width=1\textwidth,page=33]{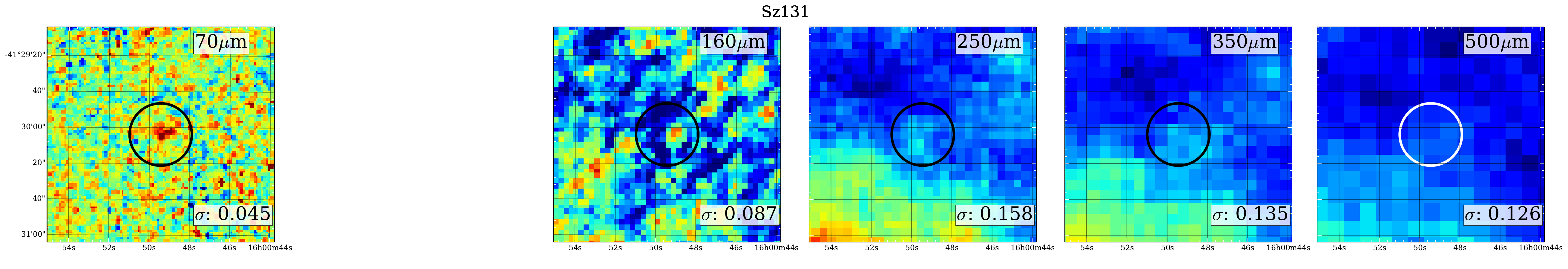}
    \includegraphics[width=1\textwidth,page=34]{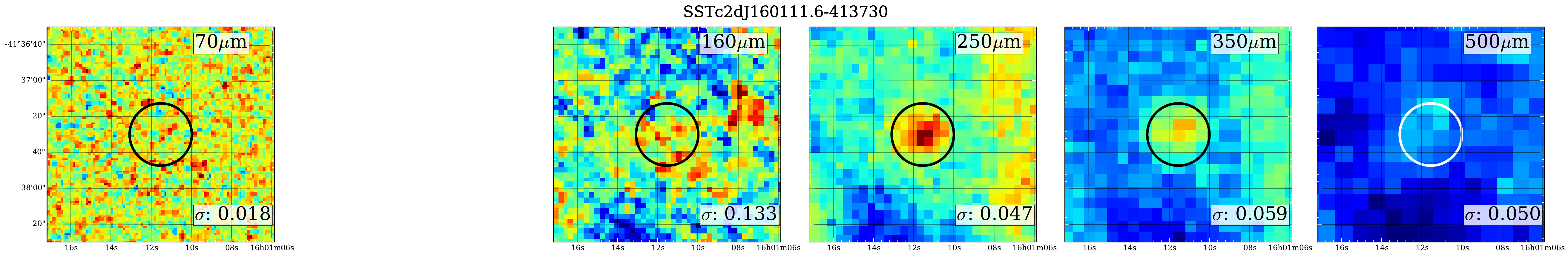}
    \includegraphics[width=1\textwidth,page=35]{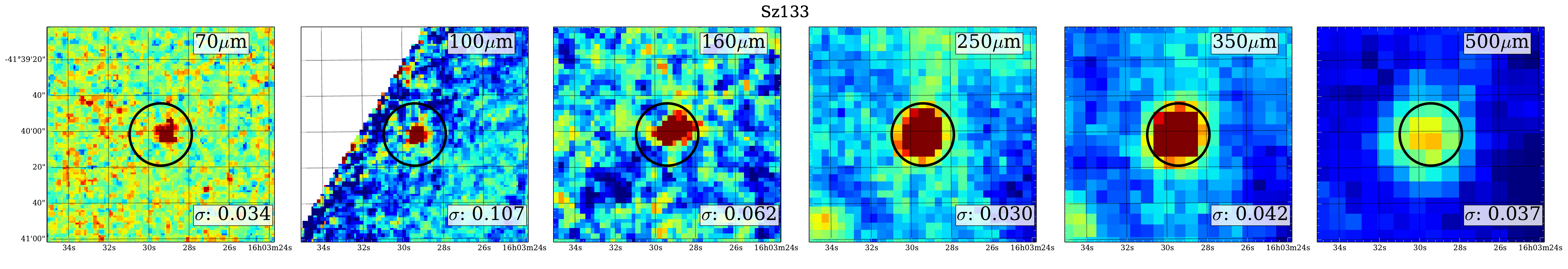}
  \label{fig:images5}
\end{figure*}

Recent studies have reported brighter 70 $\mu$m fluxes in transitional disks as compared with the median SED of "standard'' T Tauri stars in the same regions \citep{Cieza2011,Ribas2013}. To check whether this phenomenon is dependent on the conditions of certain molecular clouds, we compared the SEDs of the detected disks in Lupus with the median SED of T Tauri stars in the region. Figures \ref{fig:seds_TD}, \ref{fig:seds1} and \ref{fig:seds2} show the median SED normalized to the dereddened J-band flux for comparison, and Table \ref{tab:mediansed} shows the fluxes of this median SED.

As shown in Figure \ref{fig:seds_TD} most transitional disk candidates identified in section \ref{trans_disk_identification} show the same phenomenon: the PACS fluxes are systematically above the median SED of the T Tauri stars. Therefore, the result is confirmed so far around the isolated transitional disk T Cha \citep{Cieza2011}, in the Chamaeleon I cloud \citep{Ribas2013} and in the Lupus molecular clouds (this work).

The interpretation of this phenomenon requires detailed SED modeling with radiative transfer disk models. However, the higher PACS fluxes as compared with the median SED of the T Tauri stars suggests that the inner and outer disks of these objects follow different evolutionary paths by the time the inner gaps or holes are formed in the inner disks. Thus, the evolution of both inner and outer disk regions might not be dynamically decoupled.

\subsection{SED population analysis}
\cite{Merin2008} classified the YSOs population in Lupus by studying the shape of their SEDs, and comparing the median SED of the CTTS from Taurus with their data. They grouped the objects in four categories, based on whether their slope decay like a classical accreting optically thick disk around a low-mass star (T-Type), present infrared excess clearly weaker (L-Type) or stronger (H-Type) than the median SED or where no excess is detected (E-Type). They also introduced the category \textit{cold disks} (LU-Type), which corresponds to our criteria for transitional disks. In their work, they made use of \textit{Spitzer} and ground based photometry.

We did the same analysis for the 34 objects detected by \textit{Herschel}, but now also taking into account the photometry we extracted. For simplification, we used a different nomenclature, grouping our objects in only three categories to unveil their evolutionary state: higher infrared excess than the median values (H), indicating an early evolutionary stage; weaker (W), showing a more evolved one; and transitional candidate (T). The results can be found in Table \ref{tab:stellarparameters}, where we also compare our classification to the aforementioned work by \cite{Merin2008}. Another difference is that instead of using the Taurus median SED, we have used the Lupus one.

Apart from the transitional nature for the three new objects found in this work ($\sim$ 10\%), we find that by adding \textit{Herschel}'s photometry to the SEDs we can update the type found in \cite{Merin2008} for another 9 objects ($\sim$ 27\%). The final population for our sample is then classified as follows: $\sim$ 56\% \textit{primordial} disks (this groups T-Tauri types and stronger infrared excesses),  $\sim$ 29\% of \textit{evolved} disks and $\sim$ 15\% for transitional disk candidates. This compares well to the 67\%, 26\% and 7\% fractions for the same types of objects from the \textit{Spitzer}-only study. Table \ref{tab:stellarparameters} shows the \textit{Spitzer}-only and \textit{Spitzer}+\textit{Herschel} classifications. The relative overall fractions have not changed substantially, except on the larger fraction of transitional disk candidates.

\section {Conclusions}
\label{conclusions}
We have detected 34 objects in the Lupus cloud using \textit{Herschel} and broadened the study of the YSOs in the association. Having confirmed that our \textit{Herschel} data enable improved characterization of outer regions of protoplanetary disks, we have identified additional transitional disks in our sample. The fraction of transitional disks detected in our study, comparable in size to \cite{Ribas2013}, corrects the relatively low one from \textit{Spitzer} detections. Also, we find that all these objects have 70 $\mu$m fluxes brighter than the median SED, showing perhaps another possible intrinsic property of them. Finally, we have carried out a population analysis by studying the SED shapes of the detections, and updated the morphological type of several sources. Further studies and modeling will give us more information of these objects.

\bibliographystyle{aa}
\bibliography{biblio}

\newpage

\newpage

\begin{table*}[ht]
\centering
\caption{Class II objects detected by \textit{Herschel} in the Lupus clouds. The object types stand as higher than the median infrared excess (H), weaker (W) and transitional disk candidates (T). See text for details.}
\begin{tabular}[width=1.0\textwidth]{lc c c c c c c cl}
\hline\hline
Object & R.A.$_{J2000}$  & Dec$_{J2000}$ &  A$_{V}$ (mag) & SpT & Type (\textit{Spitzer}**) & Type (This Work) &References\tablefootmark{r}\\
\hline
\multicolumn{8}{c}{Lupus I}\\\hline
Sz65 & 15:39:27.8 & -34:46:17.1 & 1.0 & M0 & H & H &2\\
Sz66\tablefootmark{v} & 15:39:28.3 & -34:46:18.0 & 2.0 & M3 & H & H &2   \\
Sz68 & 15:45:12.9 & -34:17:30.6 & 1.5 & K2 & H & H &2  \\
Sz69 & 15:45:17.4 & -34:18:28.3 & 4.0 & M1 & W & W &2  \\
\hline
\multicolumn{8}{c}{Lupus III}\\\hline
SSTc2dJ160703.9-391112 & 16:07:03.8 & -39:11:11.3 & 2.2 & M5.5  & - & H &1  \\
Sz90 & 16:07:10.1 & -39:11:03.3 & 5.0\tablefootmark{*} & K8   & W & H &2, 5   \\
Sz91 & 16:07:11.6 & -39:03:47.5 & 1.0 & M0.5 & T & T &2 \\
Sz95 & 16:07:52.3 & -38:58:05.9& 1.0 & M1.5 & W & W &2\\
Sz96 & 16:08:12.6 & -39:08:33.5 & 1.4 & M2   & H &W &1   \\
2MASSJ1608149 & 16:08:14.9 & -38:57:14.6 & 7.0\tablefootmark{*} & M4.7   & H & T & 3, 5 \\
Sz98 & 16:08:22.5 & -39:04:46.5 & 2.5 & K5    & H & H &1  \\
Sz100 & 16:08:25.8 & -39:06:01.2 & 1.2 & M4.5  & H & H &1   \\
Sz102 & 16:08:29.7 & -39:03:11.0 & 2.5 & K2  & - &H &1   \\
Sz103 & 16:08:30.3 & -39:06:11.2 & 1.0 & M4.5  & H & H &1   \\
Sz104 & 16:08:30.8 & -39:05:48.9 & 0.7 & M5.5  & H & H &1   \\
HR5999 & 16:08:34.3 & -39:06:18.2 & 1.0\tablefootmark{*} & A7    &  H & W &2, 5  \\
RXJ1608.6-3922 & 16:08:36.2 & -39:23:02.5 & 3.0\tablefootmark{*} & K6   & - &H & 4, 5  \\
Sz108B & 16:08:42.9 & -39:06:14.4 & 0.8 & M5.5  & W & W &1  \\
Sz110 & 16:08:51.6 & -39:03:17.7 & 0.2 & M3   & H & H &1   \\
Sz111 & 16:08:54.6 & -39:37:43.1 & 0.0 & M1.5   & T & T &2 \\
Sz112\tablefootmark{j} & 16:08:55.5 & -39:02:33.9 & 1.0 & M4    & W & W &2  \\
SSTc2dJ160901.4-392512 & 16:09:01.4 & -39:25:11.9 & 0.6 & M3.5  & H & H &1  \\
Sz114\tablefootmark{j} & 16:09:01.9 & -39:05:12.4 & 1.3 & M4   & H & H &1   \\
Sz117 & 16:09:44.4 & -39:13:30.1 & 1.0 & M2     & H & W &2 \\
Sz118\tablefootmark{j} & 16:09:48.7 & -39:11:16.9 & 2.6 & K7   & H & H &1   \\
SSTc2dJ161029.6-392215 & 16:10:29.6 & -39:22:14.5 & 0.0 & M5.5  & H & T &1  \\
Sz123 & 16:10:51.6 & -38:53:13.8 & 0.0 & M3   & W & T &2   \\
\hline
\multicolumn{8}{c}{Lupus IV}\\\hline
SSTc2dJ160002.4-422216 & 16:00:02.4 & -42:22:14.6 & 2.0\tablefootmark{*} & M3.5  & H & H &1, 5  \\
IRAS15567-4141\tablefootmark{j} & 16:00:07.4 & -41:49:48.4 & 2.0 & M6.5  & W &W &1 \\
Sz130\tablefootmark{j} & 16:00:31.1 & -41:43:36.9 & 0.60 & M2   & H & W &1   \\
F403\tablefootmark{j} & 16:00:44.5 & -41:55:31.0 & 2.0 & K0    & W & H &2  \\
Sz131\tablefootmark{j} & 16:00:49.4 & -41:30:03.9 & 0.0 & M2    & H & H &2  \\
SSTc2dJ160111.6-413730\tablefootmark{v} & 16:01:11.5 & -41:37:29.9 & 0.0 & M8.5  & W &W &1  \\
Sz133\tablefootmark{j} & 16:03:29.4 & -41:40:01.8 & 10.0\tablefootmark{*} & K2    & H & H &1  \\
\hline
\label{tab:stellarparameters}
\end{tabular}
\tablefoot{
\tablefoottext{r}{References for the spectral type and A$_{V}$: (1) \cite{Mortier2011},
  (2) \cite{Comeron2008}, (3) \cite{Allen2007}, (4) \cite{Krautter1997} and (5) this work.} \\
\tablefoottext{j}{The objects marked with \textit{j} presented potential small extended emission at J band. Nevertheless, all of them have spectroscopic confirmation of being young stellar objects.}\\
\tablefoottext{v}{These objects were  missed by our detection method and were detected in the visual inspection.}\\
\tablefoottext{*}{Extinction values denoted with an asterisk indicate that were computed in this work either because they were not available in the literature or the ones found did not provide appropriate fits to the model photospheres.} \\
\tablefoottext{**}{\cite{Merin2008}.}\\
}
\end{table*}

\newpage

\begin{table*}[ht]
\centering
\caption{\textit{Herschel} photometry for the 34 YSOs detected in the maps. When no source was detected at a certain band, an upper limit is provided.}
\begin{tabular}[width=1\textwidth]{lc c c c c c cl}
\hline\hline
Object & F$_{70\mu m}$ (Jy) & F$_{100\mu m}$ (Jy) & F$_{160\mu m}$ (Jy) & F$_{250\mu m}$ (Jy) & F$_{350\mu m}$ (Jy) & F$_{500\mu m}$ (Jy) \\ 
\hline
\hline
\multicolumn{7}{c}{Lupus I}\\
\hline
Sz65& 0.47$\pm$0.12& - &0.55$\pm$0.14\tablefootmark{a}&0.77$\pm$0.15\tablefootmark{a}&0.95$\pm$0.19\tablefootmark{a}&$<$1.59 \\ 
Sz66& 0.45$\pm$0.11\tablefootmark{a}& - &0.55$\pm$0.14\tablefootmark{a}&0.81$\pm$0.16\tablefootmark{a}&0.98$\pm$0.20\tablefootmark{a}&$<$1.63 \\ 
Sz68& 3.01$\pm$0.75\tablefootmark{a}&3.80$\pm$0.95\tablefootmark{a}&11.77$\pm$2.94\tablefootmark{a}&13.06$\pm$2.61\tablefootmark{a}&10.22$\pm$2.04\tablefootmark{a}&6.74$\pm$1.35\tablefootmark{a} \\ 
Sz69& 0.12$\pm$0.03&0.19$\pm$0.05&$<$0.31&$<$1.59&$<$0.88&$<$2.13 \\ 
\hline
\hline
\multicolumn{7}{c}{Lupus III}\\
\hline
SSTc2dJ160703.9-391112& 0.07$\pm$0.02\tablefootmark{a}&0.10$\pm$0.02&0.19$\pm$0.05\tablefootmark{a}&$<$0.33&$<$0.44&$<$0.43 \\ 
Sz90& 0.36$\pm$0.09&0.37$\pm$0.09\tablefootmark{a}&0.26$\pm$0.07\tablefootmark{a}&0.51$\pm$0.10\tablefootmark{a}&0.50$\pm$0.10\tablefootmark{a}&$<$0.61 \\ 
Sz91& 0.51$\pm$0.13&0.68$\pm$0.17\tablefootmark{a}&0.72$\pm$0.18\tablefootmark{a}&0.86$\pm$0.17\tablefootmark{a}&0.62$\pm$0.12\tablefootmark{a}&0.38$\pm$0.08\tablefootmark{a} \\ 
Sz95& $<$0.05&0.05$\pm$0.01\tablefootmark{a}&0.09$\pm$0.02&$<$0.28&$<$0.30&$<$0.25 \\ 
Sz96& 0.14$\pm$0.03\tablefootmark{a}&0.12$\pm$0.03&0.10$\pm$0.02&$<$1.49&$<$1.87&$<$1.96 \\ 
2MASSJ1608149& $<$0.08&0.07$\pm$0.02\tablefootmark{a}&$<$0.11&$<$0.48&$<$0.25&$<$0.21 \\ 
Sz98& 0.64$\pm$0.16\tablefootmark{a}&0.58$\pm$0.14\tablefootmark{a}&0.50$\pm$0.12\tablefootmark{a}&0.93$\pm$0.19\tablefootmark{a}&1.14$\pm$0.23\tablefootmark{a}&$<$2.12 \\ 
Sz100& 0.17$\pm$0.04&0.23$\pm$0.06\tablefootmark{a}&0.29$\pm$0.07\tablefootmark{a}&0.59$\pm$0.12\tablefootmark{a}&$<$1.99&$<$1.33 \\ 
Sz102& 0.34$\pm$0.09&0.37$\pm$0.09\tablefootmark{a}&0.22$\pm$0.06\tablefootmark{a}&$<$3.25&$<$4.07&$<$4.37 \\ 
Sz103& 0.11$\pm$0.03\tablefootmark{a}&0.16$\pm$0.04\tablefootmark{a}&$<$0.23&$<$2.13&$<$1.64&$<$1.20 \\ 
Sz104& 0.10$\pm$0.02\tablefootmark{a}&0.18$\pm$0.04\tablefootmark{a}&$<$0.19&$<$2.08&$<$1.97&$<$2.13 \\ 
HR5999& 4.25$\pm$1.06&3.20$\pm$0.80&1.60$\pm$0.40\tablefootmark{a}&0.92$\pm$0.18\tablefootmark{a}&0.47$\pm$0.09\tablefootmark{a}&$<$3.26 \\ 
RXJ1608.6-3922& 0.40$\pm$0.10\tablefootmark{a}&0.69$\pm$0.17\tablefootmark{a}&1.36$\pm$0.34\tablefootmark{a}&2.55$\pm$0.51\tablefootmark{a}&2.83$\pm$0.57\tablefootmark{a}&2.40$\pm$0.48\tablefootmark{a} \\ 
Sz108B& 0.20$\pm$0.05\tablefootmark{a}&0.21$\pm$0.05\tablefootmark{a}&$<$0.61&$<$5.81&$<$9.68&$<$8.85 \\ 
Sz110& 0.13$\pm$0.03\tablefootmark{a}&0.09$\pm$0.02\tablefootmark{a}&$<$0.26&$<$0.57&$<$0.48&$<$0.76 \\ 
Sz111& 1.16$\pm$0.29\tablefootmark{a}& - &1.73$\pm$0.43&1.31$\pm$0.26&0.94$\pm$0.19&0.58$\pm$0.12\tablefootmark{a} \\ 
Sz112& 0.10$\pm$0.03\tablefootmark{a}&0.04$\pm$0.01&$<$0.16&$<$0.33&$<$0.27&$<$0.38 \\ 
SSTc2dJ160901.4-392512& 0.08$\pm$0.02\tablefootmark{a}&0.09$\pm$0.02\tablefootmark{a}&0.12$\pm$0.03\tablefootmark{a}&0.15$\pm$0.03\tablefootmark{a}&0.13$\pm$0.03\tablefootmark{a}&0.07$\pm$0.01\tablefootmark{a} \\ 
Sz114 &0.28$\pm$0.07&0.26$\pm$0.06\tablefootmark{a}&0.13$\pm$0.03\tablefootmark{a}&$<$0.53&$<$0.51&$<$1.11 \\ 
Sz117& 0.06$\pm$0.02\tablefootmark{a}&0.11$\pm$0.03\tablefootmark{a}&0.17$\pm$0.04\tablefootmark{a}&$<$0.82&$<$0.70&$<$0.65 \\ 
Sz118& 0.35$\pm$0.09&0.46$\pm$0.11&0.37$\pm$0.09&0.49$\pm$0.10\tablefootmark{a}&0.56$\pm$0.11\tablefootmark{a}&$<$0.98 \\ 
SSTc2dJ161029.6-392215& 0.07$\pm$0.02\tablefootmark{a}&0.10$\pm$0.02\tablefootmark{a}&$<$0.13&$<$0.29&$<$0.27&$<$0.26 \\ 
Sz123& 0.27$\pm$0.07\tablefootmark{a}&0.37$\pm$0.09\tablefootmark{a}&0.45$\pm$0.11&0.17$\pm$0.03\tablefootmark{a}&0.12$\pm$0.02\tablefootmark{a}&0.08$\pm$0.02\tablefootmark{a} \\ 
\hline
\hline
\multicolumn{7}{c}{Lupus IV}\\
\hline
SSTc2dJ160002.4-422216& 0.10$\pm$0.02\tablefootmark{a}&0.20$\pm$0.05\tablefootmark{a}&0.21$\pm$0.05\tablefootmark{a}&0.22$\pm$0.04&0.23$\pm$0.05\tablefootmark{a}&0.17$\pm$0.03\tablefootmark{a} \\ 
IRAS15567-4141& 0.11$\pm$0.03\tablefootmark{a}&0.05$\pm$0.01\tablefootmark{a}&$<$0.11&$<$0.38&$<$0.34&$<$0.22 \\ 
Sz130& 0.11$\pm$0.03&0.12$\pm$0.03\tablefootmark{a}&0.05$\pm$0.01\tablefootmark{a}&$<$0.39&$<$0.34&$<$0.34 \\ 
F403& 1.04$\pm$0.26\tablefootmark{a}&1.26$\pm$0.31\tablefootmark{a}&1.57$\pm$0.39\tablefootmark{a}&1.33$\pm$0.27\tablefootmark{a}&0.93$\pm$0.19\tablefootmark{a}&0.58$\pm$0.12\tablefootmark{a} \\ 
Sz131& 0.11$\pm$0.03& - &$<$0.10&$<$0.24&$<$0.35&$<$0.25 \\ 
SSTc2dJ160111.6-413730& $<$0.05& - &$<$0.09&0.11$\pm$0.02\tablefootmark{a}&0.07$\pm$0.01&$<$0.16 \\ 
Sz133& 0.21$\pm$0.05\tablefootmark{a}&0.38$\pm$0.10&0.32$\pm$0.08&0.37$\pm$0.07\tablefootmark{a}&0.30$\pm$0.06\tablefootmark{a}&0.20$\pm$0.04\tablefootmark{a} \\ 
\hline
\label{tab:table2}
\end{tabular}
\end{table*}

\newpage

\begin{table*}[ht]
\centering
\caption{Normalized flux densities of the median SED, lower SED (first quartile) and upper SED (fourth quartile) of the Class II objects in Lupus I, III and IV. The number of detections in each band is also included. The transitional disk candidates are excluded from this count. All values are normalized to J band.}
\begin{tabular}[width=1.0\textwidth]{lc c c c cl}
\hline\hline
Band($\mu$m) & Median &  First Quartile & Fourth Quartile & Detections\\
\multicolumn{5}{c}{(F$_{\lambda}$ arbitrary units)}\\\hline \\ 
J 1.25 & 1.000  & 1.000 & 1.000 & 28 \\ 
H 1.62 & 1.048  & 0.982 & 1.186 & 28 \\ 
K 2.15 & 0.761  & 0.673 & 1.015 & 28 \\ 
IRAC 3.6 & 0.331  & 0.260 & 0.501 & 23 \\ 
IRAC 4.5 & 0.259  & 0.182 & 0.385 & 24 \\ 
IRAC 5.8 & 0.193  & 0.143 & 0.284 & 24 \\ 
IRAC 8.0 & 0.167  & 0.111 & 0.223 & 25 \\ 
MIPS 23.68 & 0.090  & 0.054 & 0.122 & 27 \\ 
PACS 70 & 0.049  & 0.025 & 0.074 & 25 \\ 
PACS 100 & 0.042  & 0.018 & 0.056 & 25 \\ 
PACS 160 & 0.028  & 0.023 & 0.066 & 18 \\ 
SPIRE 250 & 0.023  & 0.015 & 0.045 & 15 \\ 
SPIRE 350 & 0.017  & 0.013 & 0.036 & 11 \\ 
SPIRE 500 & 0.011  & 0.007 & 0.019 & 6 \\ 
\hline
\label{tab:mediansed}
\end{tabular}
\end{table*}

\newpage

\newpage


\end{document}